\documentclass[showpacs,preprint,prb,eqsecnum,amsmath,amssymb]{revtex4}
\usepackage{graphicx}
\usepackage{dcolumn}
\usepackage{bm}
\begin{document}
\title{Opening of a pseudogap in a quasi-two dimensional superconductor
due to critical thermal fluctuations}
\author{Fusayoshi J. Ohkawa}
\affiliation{
Department of Physics, Graduate School of
Science,  Hokkaido University, Sapporo 060-0810, Japan}
\email{fohkawa@phys.sci.hokudai.ac.jp}
%
\received{6 March 2006}
\begin{abstract} 
We examine the role of the anisotropy of superconducting 
critical thermal fluctuations in the opening of a pseudogap
in a quasi-two dimensional  superconductor such as a 
cuprate-oxide high-temperature superconductor. 
When   the anisotropy  between planes and 
their perpendicular axis  is large enough and its superconducting
critical temperature  $T_c$  is high enough,
the fluctuations are much developed in its critical region so that
lifetime widths of quasiparticles are large 
and the energy dependence of the selfenergy
deviates from that of Landau's normal Fermi liquids.
A pseudogap opens in  such a critical region because
quasiparticle  spectra around the chemical potential
are swept away  due to the large lifetime widths.
The pseudogap never smoothly evolves into   
a superconducting  gap; it starts to open at a temperature
higher than $T_c$ while the superconducting  gap starts to open just  at $T_c$. 
When $T_c$ is rather low but the ratio of $\varepsilon_G(0)/k_BT_c$,
with $\varepsilon_G(0)$  the superconducting gap at $T=0$~K and
$k_B$ the Boltzmann constant, is much larger
than a value about 4 according to the mean-field theory,  
the pseudogap must be closing as temperature $T$ approaches to the low $T_c$
because thermal fluctuations become less developed as $T$ decreases.
Critical  thermal fluctuations cannot cause
the opening of a prominent  pseudogap in an almost isotropic three dimensional  
superconductor,  even if its $T_c$ is high.
\end{abstract}
\pacs{74.20.-z, 74.90.+n, 71.10.-w}
\maketitle

\section{Introduction}
\label{Introduction}

The elucidation of the mechanism
of high critical-temperature (high-$T_c$) superconductivity
occurring in cuprate-oxide  superconductors
is an important  and long standing issue  since its discovery in 1986. \cite{bednorz}
On the other hand, 
many unconventional normal-state properties are observed:
the so called spin-gap behavior or
the reduction of the nuclear magnetic relaxation (NMR) rate
with decreasing temperatures $T$, \cite{spingap} 
the opening of  a pseudogap in  quasiparticle spectra, 
\cite{Ding,Shen2,Shen3,Ino,Renner,Ido1,Ido2,Ekino} and so on.  
The issue  on the mechanism of high-$T_c$ superconductivity
cannot be settled unless not only
high-$T_c$ superconductivity but also  such unconventional  properties
are explained within a theoretical framework.
It is widely believed that
the reduction of the NMR rate above $T_c$ is due to the
opening of the pseudogap.  
It is a key issue to clarify the relation between  the pseudogap  
 above $T_c$ and a superconducting  (SC) gap   below $T_c$
or whether  or not  the pseudogap smoothly evolves into the SC gap.

One may argue that the opening of a pseudogap must be 
a precursor effect of a possible 
low-temperature instability, 
antiferromagnetism, superconductivity, or an exotic one. 
Another may argue that a pseudogap must open due to  
fluctuations corresponding to one or some of 
antiferromagnetism, superconductivity, and  exotic ones.
In actual,  many possible mechanisms of pseudogaps have already
been proposed along these scenarios.
\cite{ps-randeria,ps-narikiyo,ps-deisz,ps-dagotto,ps-yamada,ps-tesanovic,%
ps-loktev,ps-monthoux,ps-lee,ps-sunko,ps-ramsak,ps-mamedov,%
ps-marchetti,ps-mitsen,ps-metzner,ps-gu,ps-kampf}
Since cuprate-oxide superconductors are highly anisotropic
quasi-two dimensional ones, it is also a reasonable argument
that, even if either of these scenarios is relevant, low dimensionality
must play a crucial role.
If a second-order phase transition occurred at a non-zero  
critical temperature $T_c$  in one or two dimensions, 
eventual  effects of critical thermal  fluctuations 
or their integrated effects over the wave-number space
would diverge at the nonzero $T_c$.
This leads to a conclusion that no order is possible
at non-zero temperatures in one and two dimensions. \cite{Mermin}
It also leads to a speculation that $T_c$ of highly anisotropic
quasi-low dimensional superconductors  must  be substantially
reduced by SC critical thermal fluctuations and the reduction of 
$T_c$ must be accompanied by some normal-state anomalies.
Since pseudogap structures are certainly
substantial in SC critical regions of cuprate-oxide superconductors, 
it must be clarified first of all how a crucial role 
SC critical fluctuations can play in the reduction of $T_c$
and the opening of pseudogaps,  even if any other type of fluctuations  
or any other mechanism plays a  role or a major role.

Critical temperatures $T_c$ are reduced by SC critical
thermal fluctuations. On the other hand,  SC gaps at $T=0$~K, 
$\varepsilon_G(0)$, can never be reduced by them
because  they vanish at $T=0$~K.
Therefore, large ratios  of  $\varepsilon_G(0)/k_BT_c \agt 8$, 
\cite{Renner,Ido1,Ido2}
with  $k_B$ the Boltzmann constant, are pieces of evidence that $T_c$
are actually substantially reduced even if observed $T_c$ are high.
 It is plausible that
the opening of a pseudogap must be one of normal-state anomalies
accompanying the reduction of $T_c$. 

It has  been shown in a previous paper \cite{pseudogap}
that when  correlation lengths of
SC fluctuations are long enough at high enough temperatures
 in complete two dimensions
the renormalization of quasiparticles due to SC fluctuations can
cause the opening of a pseudogap.
One of the main purposes of this paper is to  
clarify the role of the anisotropy of SC critical  thermal fluctuations
in  the renormalization of quasiparticles and
the opening of a pseudogap in quasi-two dimensions.
When we consider cuprate-oxide superconductors,
we should take a repulsive strong-coupling model.
However, we consider an attractive intermediate-coupling model 
\cite{ps-deisz,ps-dagotto}
in order to demonstrate the essence of a  mechanism proposed
in this paper.
This paper is organized as follows.
The formulation  is presented in Sec.~\ref{SecFormulation}.
It is demonstrated in Sec.~\ref{SecOpening} that 
a pseudogap can open
because of highly anisotropic SC critical thermal fluctuations.
It is argued  in Sec.~\ref{SecApplication}
that the fluctuations
must play a role in the opening of pseudogaps on cuprate-oxide superconductors.
Discussion is  given in Sec.~\ref{SecDiscussion}.
Conclusion is presented  in Sec.~\ref{SecConclusion}.
An argument is presented in Appendix~\ref{SecAppendix} 
in order to show the relevance of a scenario that high-$T_c$ superconductivity of 
cuprate oxides occurs 
in an attractive intermediate-coupling regime for superconductivity,
which is realized in a repulsive strong-coupling regime for electron correlations.

\section{Formulation}
\label{SecFormulation}

We consider  an attractive intermediate-coupling
model on a quasi-two dimensional lattice composed of square lattices:
\begin{equation}\label{EqModel}
{\cal H} = \sum_{ij\sigma} t_{ij} a_{i\sigma}^\dag a_{j\sigma}
+ \frac1{2} 
\sum_{ij \sigma\sigma^\prime} U_{ij}  
a_{i\sigma}^\dag a_{j\sigma^\prime} ^\dag 
a_{j\sigma^\prime} a_{i\sigma}  .
\end{equation}
When transfer integrals $t_{ij}$ between
nearest and next nearest neighbors on a plane,  $-t$ and $-t^\prime$,
are considered, the dispersion relation of electrons is given by
\begin{equation}
E({\bf k}) = - 2 t \left[\cos(k_xa)+\cos(k_ya) \right]
- 4 t^\prime \cos(k_xa)\cos(k_ya) ,
\end{equation}
with $a$ the lattice constant of square lattices;
the bandwidth is $8|t|$.
We denote attractive interactions $U_{ij}$ between onsite  and
nearest-neighbor pairs on a plane
by $U_0$ and $U_1$. We consider two  models:
(i) $U_0/|t | \simeq -4$ and $U_1 = 0$,  and 
(ii) $U_0 =0$ and $U_1/|t| \simeq -4$.
In model (i), SC fluctuations corresponding to
isotropic $s$-wave  superconductivity are developed.
In model (ii), 
SC fluctuations corresponding to
anisotropic $s$-wave, $p$-wave, or  $d_{x^2-y^2}$- or
$d\gamma$-wave superconductivity can be developed.
We consider only two cases: the isotropic $s$-wave  case in model (i) and 
the $d\gamma$-wave  case in model (ii). 
We  consider
quasi-two dimensional features phenomenologically by introducing 
an anisotropy factor for  correlation lengths of SC fluctuations, 
as is discussed below.

We define a SC susceptibility  for singlet superconductivity by \cite{couplings}
\begin{equation}\label{EqChiSC}
\chi_{\Gamma\Gamma^\prime} (i\omega_l, {\bf q}) =
\int_{0}^{1/k_BT} \hspace*{-10pt} d \tau e^{-i \omega_l \tau}
\frac1{N}\sum_{{\bf k}{\bf p}}
\eta_{\Gamma} ({\bf k}) \eta_{\Gamma^\prime} ({\bf p})  
\left<
a_{{\bf k}+\frac1{2}{\bf q}\uparrow}(\tau)
a_{-{\bf k}+\frac1{2}{\bf q}\downarrow}(\tau)
a_{-{\bf p}+\frac1{2}{\bf q}\downarrow}^\dag 
a_{{\bf p}+\frac1{2}{\bf q}\uparrow}^\dag 
\right> ,
\end{equation}
where $\omega_l=2l\pi k_BT$, with $l$ an integer, is a bosonic energy.
Here, $\eta_{\Gamma} ({\bf k})$ is a form factor of $\Gamma$-wave Cooper pairs:
\begin{equation}
\eta_{s} ({\bf k})  = 1,  
\end{equation}
for the isotropic $s$ wave of model (i) and
\begin{equation}
\eta_{\Gamma} ({\bf k})  =\left\{ \begin{array}{cl}
\cos(k_xa)+\cos(k_ya), &  \Gamma=s\\
\cos(k_xa)-\cos(k_ya), &  \Gamma=d\gamma
\end{array}\right.  ,
\end{equation}
for the anisotropic $s$ and $d\gamma$ waves of model (ii).
We assume that the conventional condensation
of Cooper pairs with zero total momenta occurs below 
a critical temperature $T_c$. When superconductivity of 
the isotropic $\Gamma=s$ or $\Gamma=d\gamma$ wave occurs,
the homogeneous and static part of the
SC susceptibility shows a divergence at $T_c$.
Superconducting $T_c$ can be determined from the condition of 
\begin{equation}\label{EqCondTc}
\Bigl[\chi_{\Gamma\Gamma} (0, |{\bf q}|\rightarrow 0) |t| 
\Bigr]_{T\rightarrow T_c+0} \rightarrow + \infty.
\end{equation}
The divergence  implies  that critical fluctuations can play a role,
at least, in SC critical regions of highly anisotropic quasi-two dimensions.

We divide the selfenergy correction  into two terms:
\begin{equation}
\Sigma_{\sigma}(i\varepsilon_n,{\bf k}) =
\Sigma_{\sigma}^{(SC)}(i\varepsilon_n,{\bf k}) 
+\Sigma_{\sigma}^{\prime}(i\varepsilon_n,{\bf k}) ,
\end{equation}
where $\varepsilon_n=(2n+1)k_B T$, with $n$ an integer, is a fermionic energy.
The first term $\Sigma_{\sigma}^{(SC)}(i\varepsilon_n,{\bf k}) $
is  due to SC fluctuations of the isotropic $s$ or $d\gamma$ wave and is of linear order in 
$\chi_{\Gamma\Gamma} (i\omega_l, {\bf q})$, 
and the second term
$\Sigma_{\sigma}^{\prime}(i\varepsilon_n,{\bf k})$
is due to other  fluctuations such as SC fluctuations of other waves, 
charge fluctuations,
higher-order terms in SC and charge fluctuations, and so on.
When fluctuations of a single wave,
the $s$ or $d\gamma$ wave, are considered, \cite{couplings}
it follows that
\begin{equation}\label{EqSigma}
\Sigma_{\sigma}^{(SC)}(i\varepsilon_n,{\bf k}) =
- \frac{k_BT}{N}\sum_{\omega_l{\bf q}}
U_{\Gamma}^2\eta_{\Gamma}^2
\left({\bf k}-\frac{1}{2}{\bf q}\right) 
\chi_{\Gamma\Gamma}(i\omega_l,{\bf q}) 
G_{-\sigma}(-i\varepsilon_n-i\omega_l,-{\bf k}-{\bf q}) ,
\end{equation}
with  $N$ the number of unit cells,
\begin{equation}
U_{\Gamma} =\left\{ \begin{array}{ll}
 U_0, &  \Gamma=s \\
  U_1,&  \Gamma=d\gamma
\end{array}\right.  ,
\end{equation}
an effective attractive interaction for the $s$ or $d\gamma$ wave, 
 and
\begin{equation}\label{EqG}
G_{\sigma}(i\varepsilon_n, {\bf k}) =
\frac1{i\varepsilon_n + \mu - E({\bf k}) - 
\Sigma_{\sigma}(i\varepsilon_n,{\bf k}) },
\end{equation}
the renormalized Green function, with
$\mu$  the chemical potential.
Since we are interested in SC critical fluctuations
of $\Gamma$-wave superconductivity,
we assume that  critical points of other instabilities
are a little far way from the  critical point or region 
of $\Gamma$-wave superconductivity.
Then, the energy and wave-number dependences of 
$\Sigma_{\sigma}^{\prime}(i\varepsilon_n,{\bf k})$
can play no significant role in the SC critical region, so that we simply assume 
\begin{equation}
\Sigma_{\sigma}^{\prime}(i\varepsilon_n,{\bf k}) =
- i \gamma \frac{\varepsilon_n}{|\varepsilon_n| }.
\end{equation}
Here,  $\gamma$ is the lifetime width of quasiparticles
due to other fluctuations except for those of the considered $s$-wave
or $d\gamma$-wave SC fluctuations.
Although it is desirable to calculate selfconsistently
the total selfenergy $\Sigma_{\sigma}(i\varepsilon_n,{\bf k}) $
to satisfy Eq.~(\ref{EqSigma}), \cite{selfconsistency}
we approximately replace as a first-order approximation
$G_{\sigma}(i\varepsilon_n, {\bf k})$ in the right-hand side
of Eq.~(\ref{EqSigma}) by an {\it unperturbed} one given by
\begin{equation}\label{EqG0}
G_{\sigma}^{(0)}(i\varepsilon_n, {\bf k}) =
\frac1{\displaystyle i\varepsilon_n +\mu -E({\bf k}) 
- \Sigma_{\sigma}^{\prime}(i\varepsilon_n,{\bf k})  } .
\end{equation}
Since critical fluctuations are restricted to a narrow region around
${\bf q}=0$,   Eq.~(\ref{EqSigma}) is approximately given by
\begin{equation}\label{EqSigma1}
\Sigma_{\sigma}^{(SC)}(i\varepsilon_n,{\bf k}) =
- U_{\Gamma}^2\eta_{\Gamma}^2 \left({\bf k}\right)
k_B T \sum_{\omega_l}
G_{-\sigma}^{(0)}(-i\varepsilon_n- i\omega_l, {\bf k}) 
\frac1{N} \sum_{|{\bf q}_\parallel|\le q_c} \sum_{q_z}
\chi_{\Gamma\Gamma}(i\omega_l,{\bf q}) ,
\end{equation}
where the summation over 
${\bf q}_\parallel=(q_x,q_y)$ is
restricted to $|{\bf q}_\parallel|\le q_c$.
The density of states  is given by
the retarded Green function in such a way that
\begin{equation}
\rho(\varepsilon) =
\frac1{N}  \sum_{\bf k}\rho_{\bf k} (\varepsilon) ,
\end{equation}
with
\begin{equation}\label{EqRho-k}
\rho_{\bf k}(\varepsilon) =
\left(  - \frac1{\pi} \right) 
\mbox{Im}   \left[  \frac1{\displaystyle 
\varepsilon +\mu -E({\bf k})  + i \gamma
- \Sigma_{\sigma}^{(SC)}(\varepsilon + i0,{\bf k}) } 
\right] ,
\end{equation}
being the spectral weight of quasiparticles with 
energy $\varepsilon$ and wave number ${\bf k}$.

When an anisotropic three dimensional model 
is considered, 
it is straightforward to carry out  selfconsistently a numerical calculation
according to the above formulation.
Since the anisotropy of SC thermal critical fluctuations plays 
the most crucial role in the opening of a pseudogap,
it is much more convenient to use a phenomenological SC susceptibility, which can
explicitly include the anisotropy factor for SC correlation lengths,
within the two-dimensional model (\ref{EqModel})
than to use a microscopically derived one for the 
anisotropic three dimensional model.
The SC retarded susceptibility can be approximately but well described by
a phenomenological one: 
\begin{equation}\label{EqPhChi}
\chi_{\Gamma\Gamma}(\omega+i0,{\bf q}) =
\frac{\chi_{\Gamma}(0) \kappa^2}
{\displaystyle 
\kappa^2 + (q_\parallel a)^2 + \delta^2 ( q_z c)^2
+ \alpha \omega 
-i \frac{\omega}{\Gamma_{SC}|t|} } .
\end{equation}
This  is 
similar to the wellknown one for the spin susceptibility except for
the existence of the so called $\omega$-linear  real term, 
$\alpha \omega$, with $\alpha$ being real. According to a  microscopic calculation,
such as is carried out in the previous paper, \cite{pseudogap}
we can show that $|\alpha| \ll 1$;
the $\omega$-linear term  is ignored because    it
plays no significant role.
We can also show  that
Eq.~(\ref{EqPhChi}) can be  used  not only 
for $|q_\parallel a| \ll 1$ and $|\omega/t| \ll 1$ 
but also for a little  larger region 
than that, that is,  at least for $|q_\parallel a| \alt 2$ and $|\omega/t| \alt 2$.
In Eq.~(\ref{EqPhChi}),  $\chi_{\Gamma}(0)$ is the static homogeneous one
or $\chi_{\Gamma}(0)=\chi_{\Gamma\Gamma}(0,|{\bf q}|\rightarrow 0) $,
and $\Gamma_{SC}|t|$ is an energy scale of SC fluctuations.
We introduce no cutoff in the $\omega$ or energy integration
of Eq.~(\ref{EqSigma1}) because Eq.~(\ref{EqSigma1})
is never sensitive of the cutoff energy $\omega_c$
 as long as $\omega_c/|t| \agt2$.
 In Eq.~(\ref{EqPhChi}), $c$ 
 is introduced for the lattice constant along the $z$ axis,
and a factor $\kappa$ and 
an anisotropy factor $\delta$ are introduced in such a way that
SC correlation lengths parallel to 
the $x$-$y$ planes and along the $z$ axis are
$a/\kappa$ and $\delta c/\kappa$, respectively.
We introduce a cut-off wave number $q_c=\pi/3a$ in the ${\bf q}_\parallel$
integration of Eq.~(\ref{EqSigma1});
the ${\bf q}$ integration  is carried out
over a region of $|{\bf q}_\parallel  | a  \le \pi/3$ and $|q_z | c\le \pi$. 
According to the definition of the anisotropy factor $\delta$ by Eq.~(\ref{EqPhChi}),
the absolute magnitudes of $a$ and $c$,   
the anisotropy of the lattice constants, or the difference between $a$ and $c$ 
plays no role in the framework of this paper. 

In general, 
$\chi_{\Gamma}(0)|t| \propto [(T-T_c)/T_c]^{-\lambda}$ and
$\kappa^2 \propto [(T-T_c)/T_c]^\lambda$ as $T \rightarrow T_c+0$,
with $\lambda$ being a critical exponent. \cite{ComLambda}
Since $\chi_{\Gamma}(0) |t| \kappa^2$
is almost independent of $T$ at  $T>T_c$ in such a way that
\begin{equation}\label{EqO1}
\chi_{\Gamma}(0) |t| \kappa^2   \simeq 1 ,
\end{equation}
%
we assume $U_{\Gamma}/|t| \simeq - 4$ or
\begin{equation}
g_{\Gamma} \equiv \frac1{\pi} \left(U_{\Gamma}/t\right)^2
 \chi_{\Gamma}(0) |t| \kappa^2 = 4 ,
\end{equation}
in both  the two cases.
We also assume $t^\prime = - 0.3 t<0$ 
for transfer integrals and
$\mu/|t|=-0.5$ for the chemical potential.

\section{Opening of a pseudogap at a critical temperature}
\label{SecOpening}

In this section, we restrict our examination to  
the SC critical temperature, 
$T=T_c$, so that $\kappa=0$.
Although $T_c$, which is determined from
Eq.~(\ref{EqCondTc}),  depends on other parameters, we treat it  as
an independent one.
Then, free parameters  are  $T_c$, 
$\delta$, $\Gamma_{SC}$,  and $\gamma$.
Qualitative features  are the same among results for different $\gamma$, 
unless $\gamma$ are extremely large such as $\gamma/|t| \gg 1$; 
if a fine structure appears in a physical property it is
sharper for a smaller $\gamma$.  \cite{ComGamma}
We present here only results for $\gamma/|t|=0.5$, that is,  
results  for $\gamma/|t|=0.5$,  and
various sets of $T_c$,  $\delta$, and $\Gamma_{SC}$.

\begin{figure*}
\hspace*{-0.5cm}
\includegraphics[width=7.0cm]{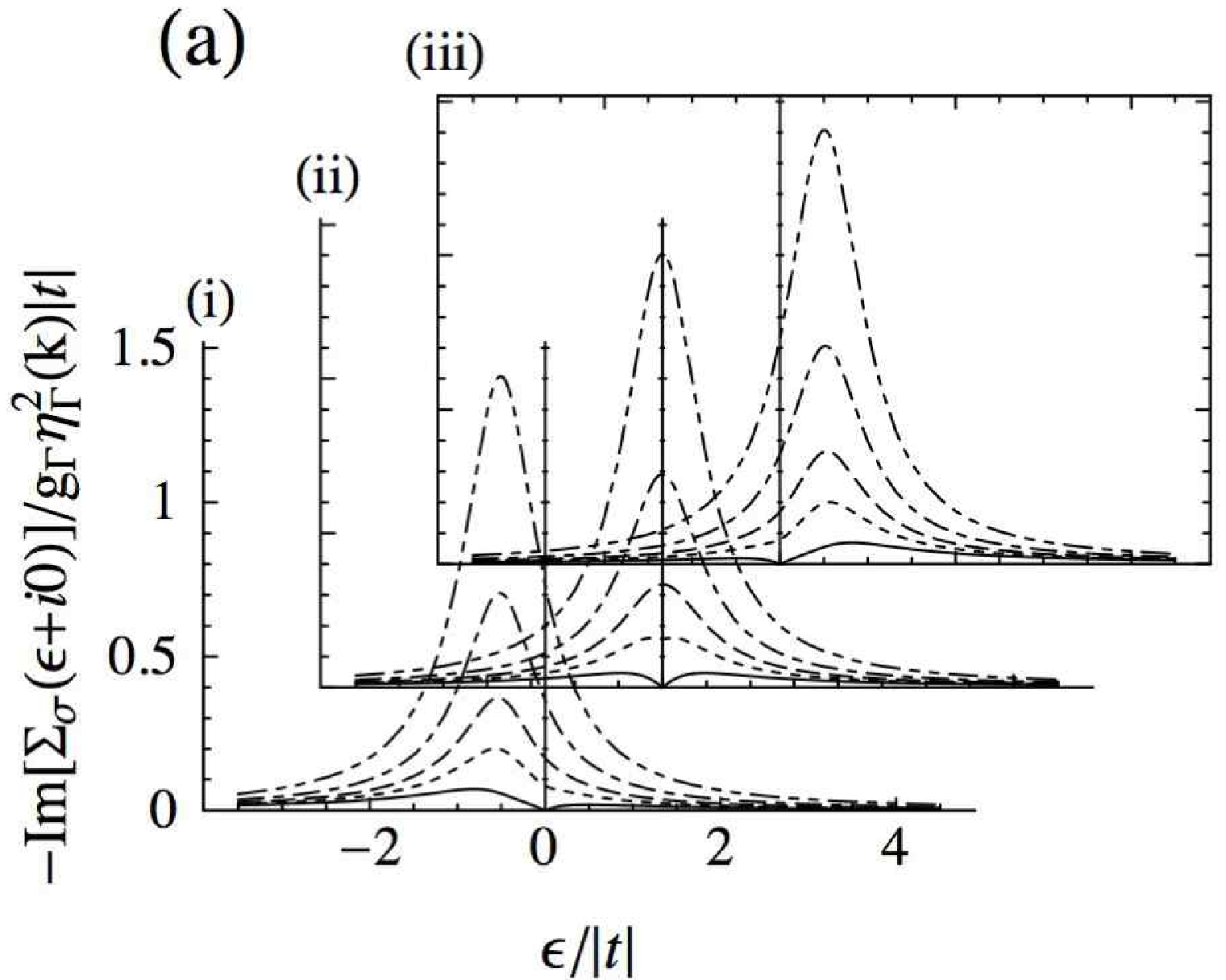}
\includegraphics[width=7.0cm]{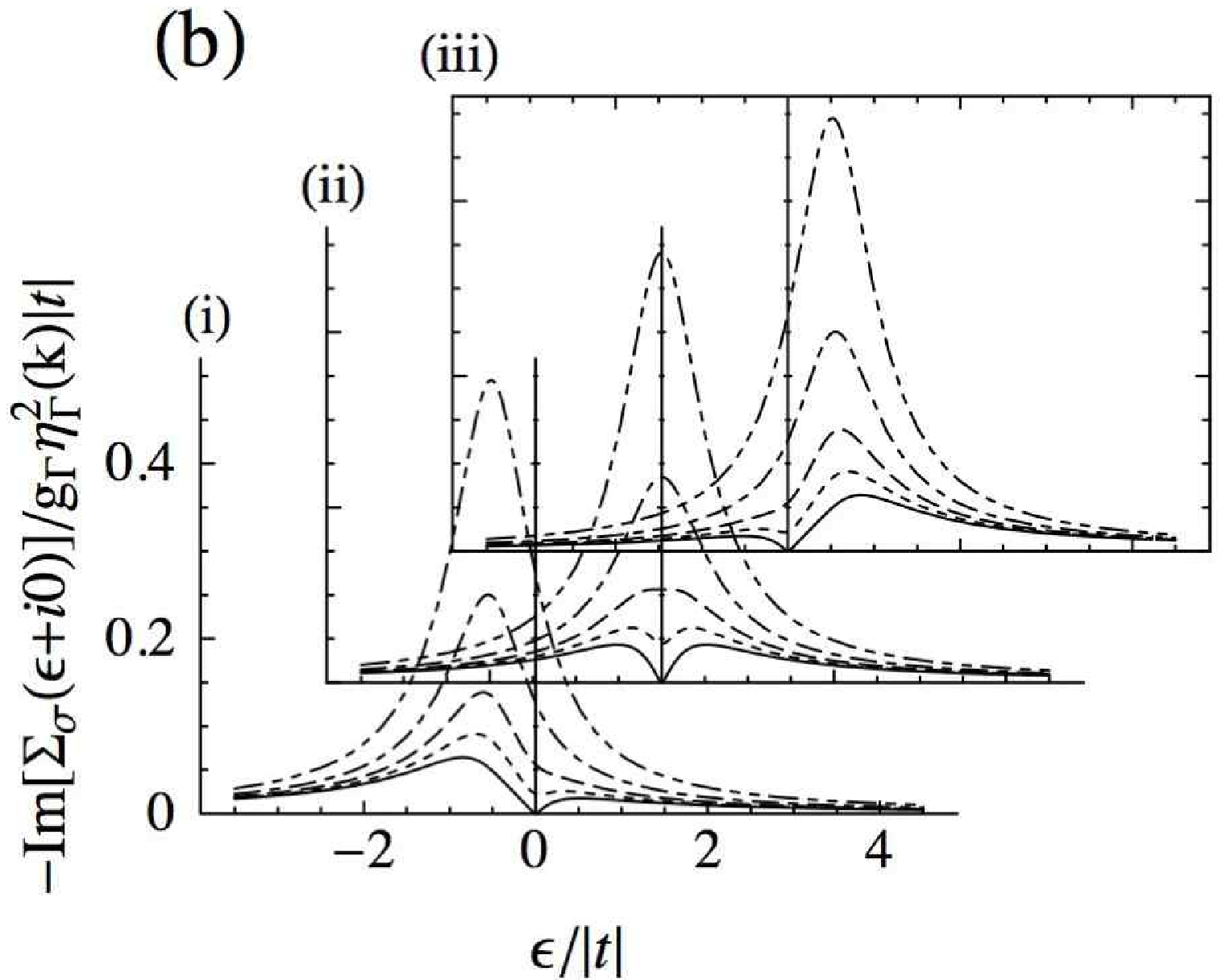}
\hspace*{-0.5cm}
\includegraphics[width=7.0cm]{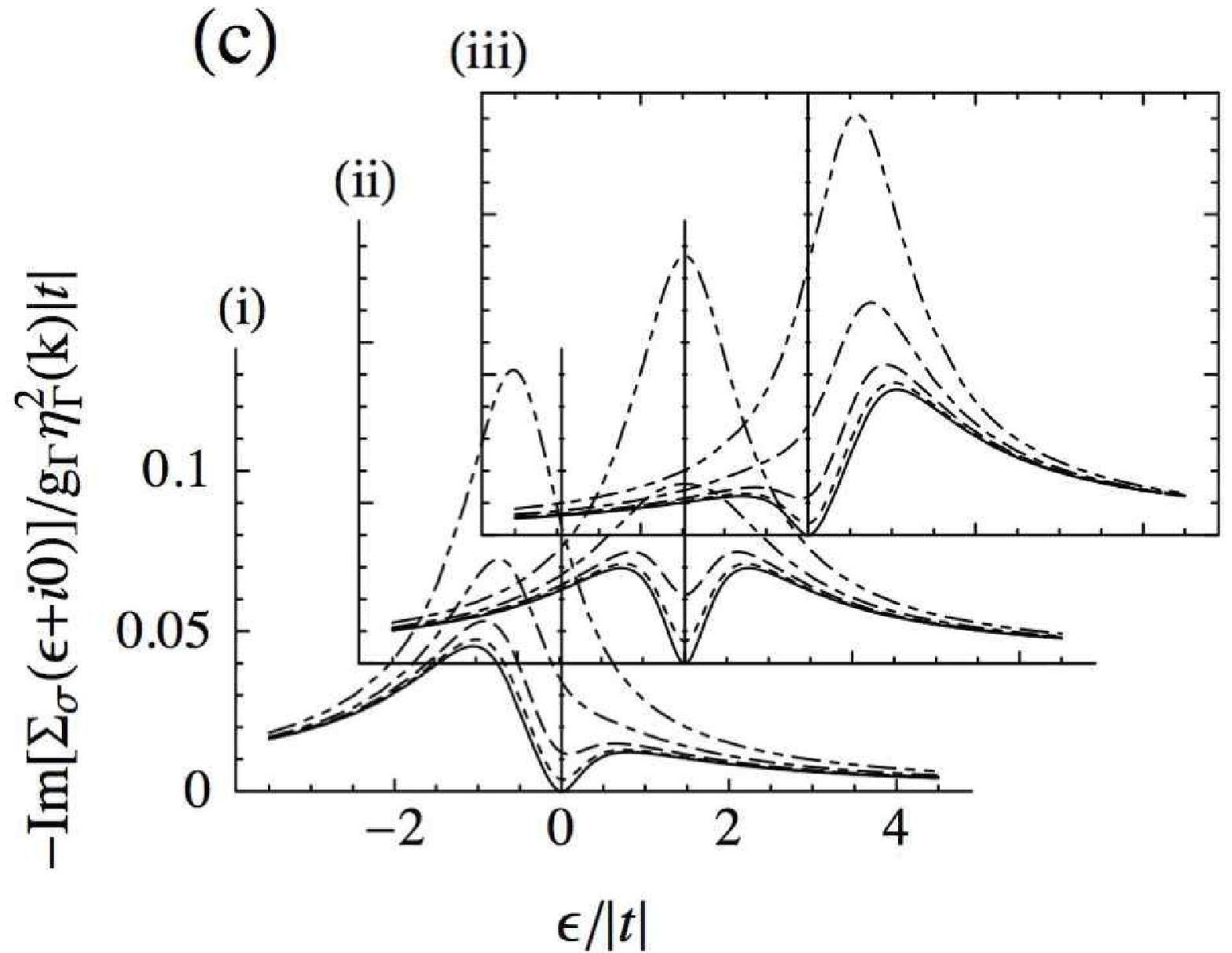} 
\includegraphics[width=7.0cm]{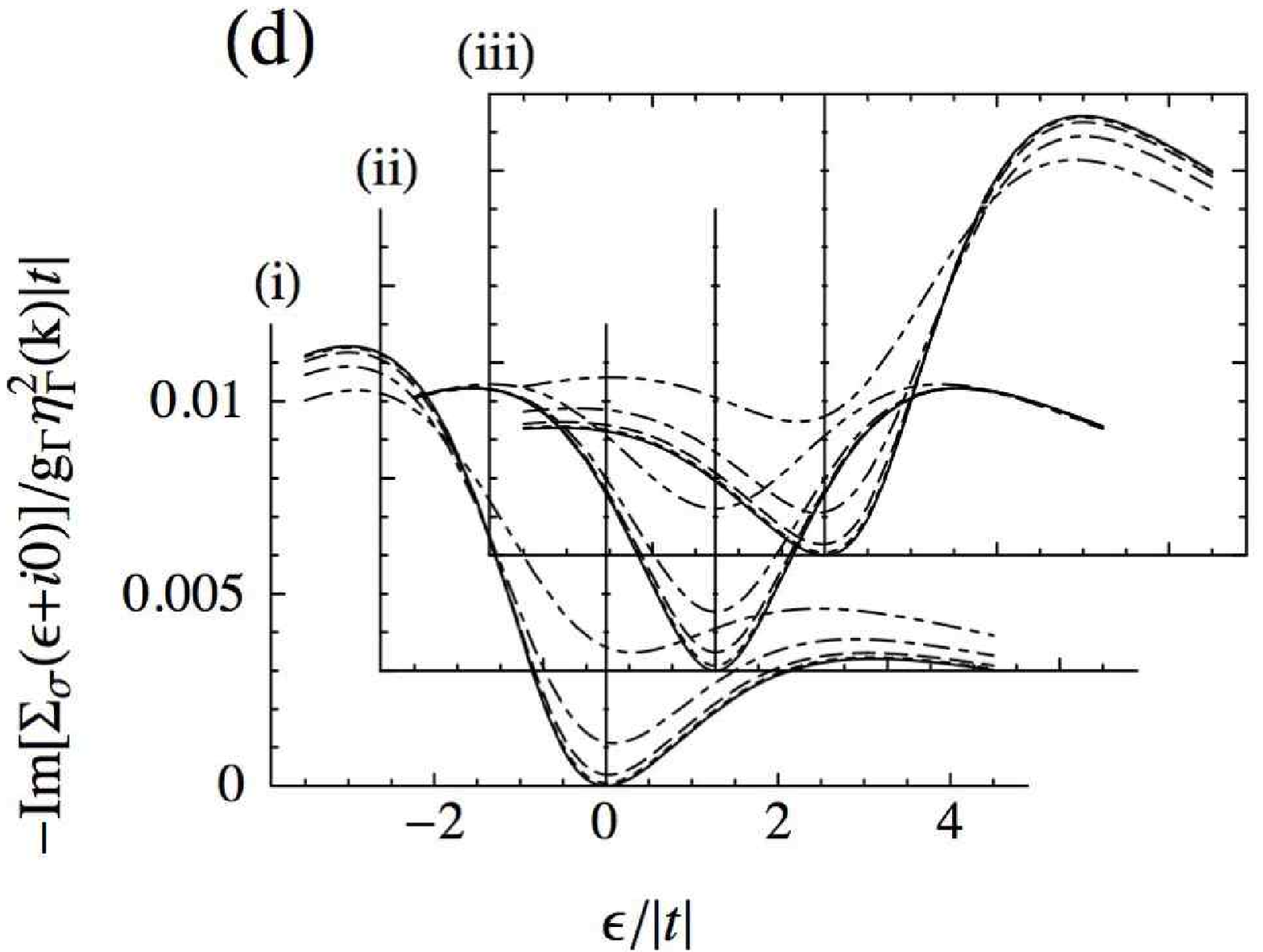}
\caption[1]{
The imaginary part of the selfenergy
for  $\kappa=0$, $\Gamma_{SC}=0.2$ and $\gamma/|t|=0.5$: 
(a) $ \delta=0.01$, (b) $\delta=0.1$,
(c) $\delta=0.3$, and (d) $\delta=1$;
(i) $E({\bf k})=\mu -0.5|t|$, 
(ii) $E({\bf k})=\mu$, and 
(iii) $E({\bf k})=\mu +0.5|t|$.
In each figure, 
solid, dashed, broken, chain, and chain double-dashed
lines show results for $k_BT_c/|t|=0$,
0.05, 0.1, 0.2, and 0.4, respectively.
The imaginary part is  larger for a smaller $\delta$ and a higher $T_c$.
When $\delta$ is small enough and $T_c$ is high enough,
the $\varepsilon$ dependence is different from that of conventional normal
Fermi liquids; there is no minimum at the zero energy or the chemical potential.
}
\label{Self}
\end{figure*}
\begin{figure}
\hspace*{-0.5cm}
\includegraphics[width=8.0cm]{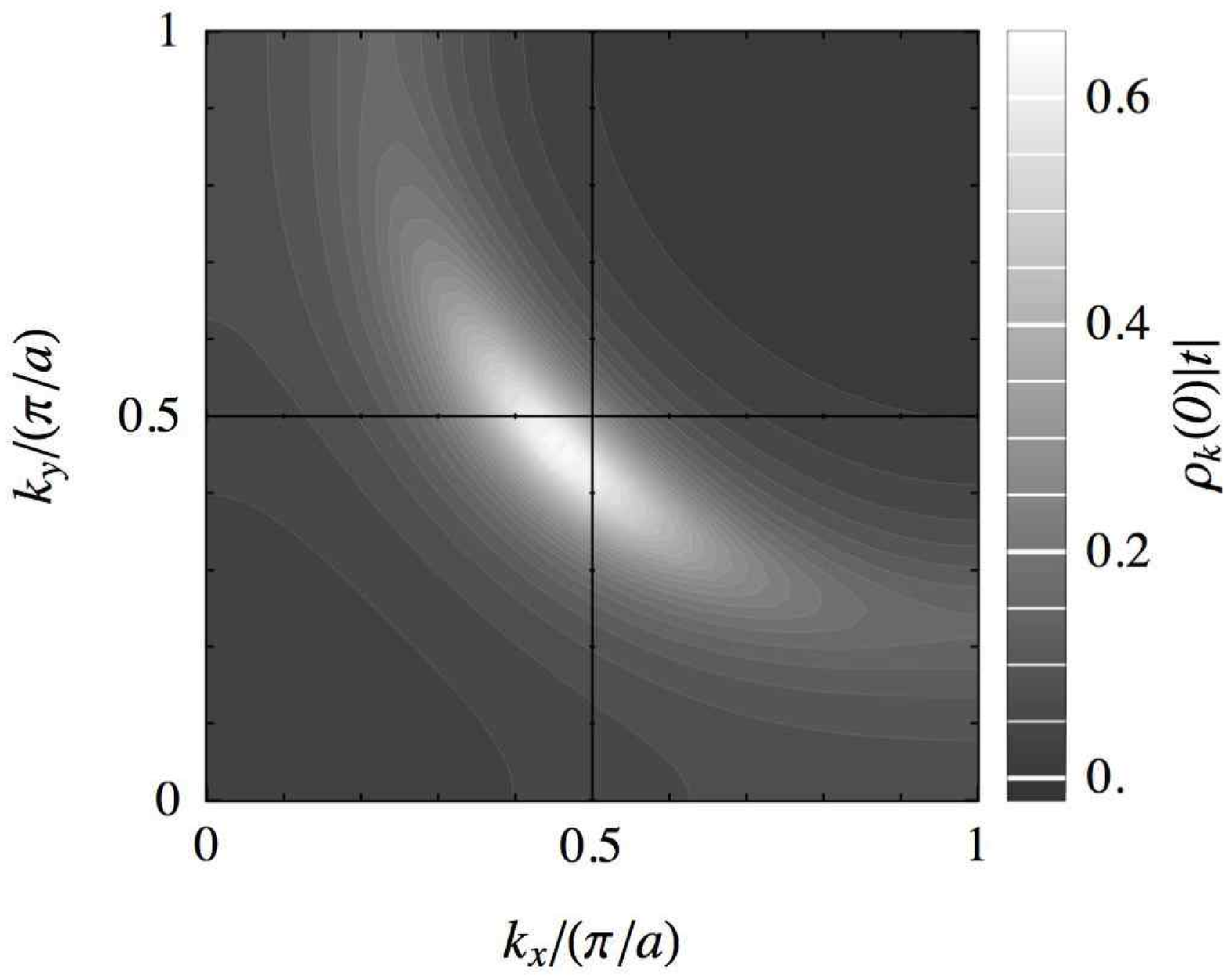}
\caption[2]{
Spectral weight $\rho_{\bf k}(\varepsilon=0)$
defined by Eq.~(\ref{EqRho-k})
 in a quarter of the two-dimensional
Brillouin zone 
for the $d\gamma$ wave,   $\kappa=0$, $k_BT_c/|t|=0.2$,
$\delta=0.1$, 
$g_{d\gamma}=4$, $\gamma/|t|=0.5$, and $\mu/|t|=-0.5$.
The spectral weight 
 is large  around ${\bf k} =(\pi/2a,\pi/2a)$ but  is small
 around ${\bf k} =(\pi/a,0)$ and ${\bf k} =(0,\pi/a)$.
}
\label{contour}
\end{figure}

Figure~\ref{Self} shows the imaginary part of the selfenergy,
\begin{equation}\label{EqImS}
- \mbox{Im}\left[\Sigma_\sigma(\varepsilon+i0, {\bf k}) \right]/
g_{\Gamma}\eta_{\Gamma}^2({\bf k}) |t| ,
\end{equation}
as a function of $\varepsilon$ for  three cases of $E({\bf k})-\mu$;
the ${\bf k}$ dependence of Eq.~(\ref{EqImS}) comes through $E({\bf k})$. 
When fluctuations are isotropic $(\delta=1)$,
the selfenergy  is small and its $\varepsilon$
dependence is consistent with that of Landau's normal Fermi liquids,
as is shown in Fig.~\ref{Self}(d).
As long as $T_c$ or $T$ is low enough, 
lifetime widths of quasiparticles  are  also small and 
the $\varepsilon$ dependence of the selfenergy 
is also consistent with that of Landau's normal Fermi liquids,
even when the anisotropy is large.
When fluctuations are anisotropic $(\delta \ll 1)$ and $T_c$ or $T$ is
high enough, on the other hand,
the imaginary part of the selfenergy is large and it has a peak
around the chemical potential, as is shown in Figs.~\ref{Self}(a), (b), and (c).
The bandwidth of quasiparticles is about $8|t|$,
$\eta_{s}^2({\bf k})=1$ for $s$ wave, 
$\eta_{d\gamma}^2(\pm\pi/a,0)=\eta_{d\gamma}^2(0,\pm\pi/a)=4$
for $d\gamma$ wave, and
we assume $g_\Gamma =4$ for both the waves.
Therefore, quasiparticles are not well defined or incoherent
on the whole Fermi surface 
in case of the $s$ wave provided that
%
$- \mbox{Im}\left[\Sigma_\sigma(\varepsilon+i0, {\bf k}) \right]/
g_{s}\eta_{s}^2({\bf k}) |t|  \agt  1$.
%
They  are incoherent around $(\pm\pi/a,0)$ and $(0,\pm\pi/a)$
in case of the $d\gamma$ wave provided that
$- \mbox{Im}\left[\Sigma_\sigma(\varepsilon+i0, {\bf k}) \right]/
g_{d\gamma}\eta_{d\gamma}^2({\bf k}) |t|  \agt  1/4$.
%
 For example, 
Fig.~\ref{contour} shows the spectral weight $\rho_{\bf k}(\varepsilon=0)$
for $d\gamma$-wave case, which is  defined by Eq.~(\ref{EqRho-k}).
The spectral weight $\rho_{\bf k}(\varepsilon=0)$
is large around ${\bf k} =(\pm\pi/2a,\pm\pi/2a)$
 so that quasiparticles are rather well defined there, while
$\rho_{\bf k}(\varepsilon=0)$ is  small
 around ${\bf k} =(\pm\pi/a,0)$ and ${\bf k} =(0,\pm\pi/a)$
 so that quasiparticles are not well defined there.

Since superconductivity can only occur when
lifetime widths are small enough, 
Fig.~\ref{Self} implies that 
the reduction of $T_c$ is large in a highly anisotropic quasi-two dimensional
superconductor even when  its observed $T_c$ is high.
The reduction of $T_c$ must be small in  an almost isotropic three dimensional one.

\begin{figure*}
\hspace*{-0.5cm}
\includegraphics[width=7.0cm]{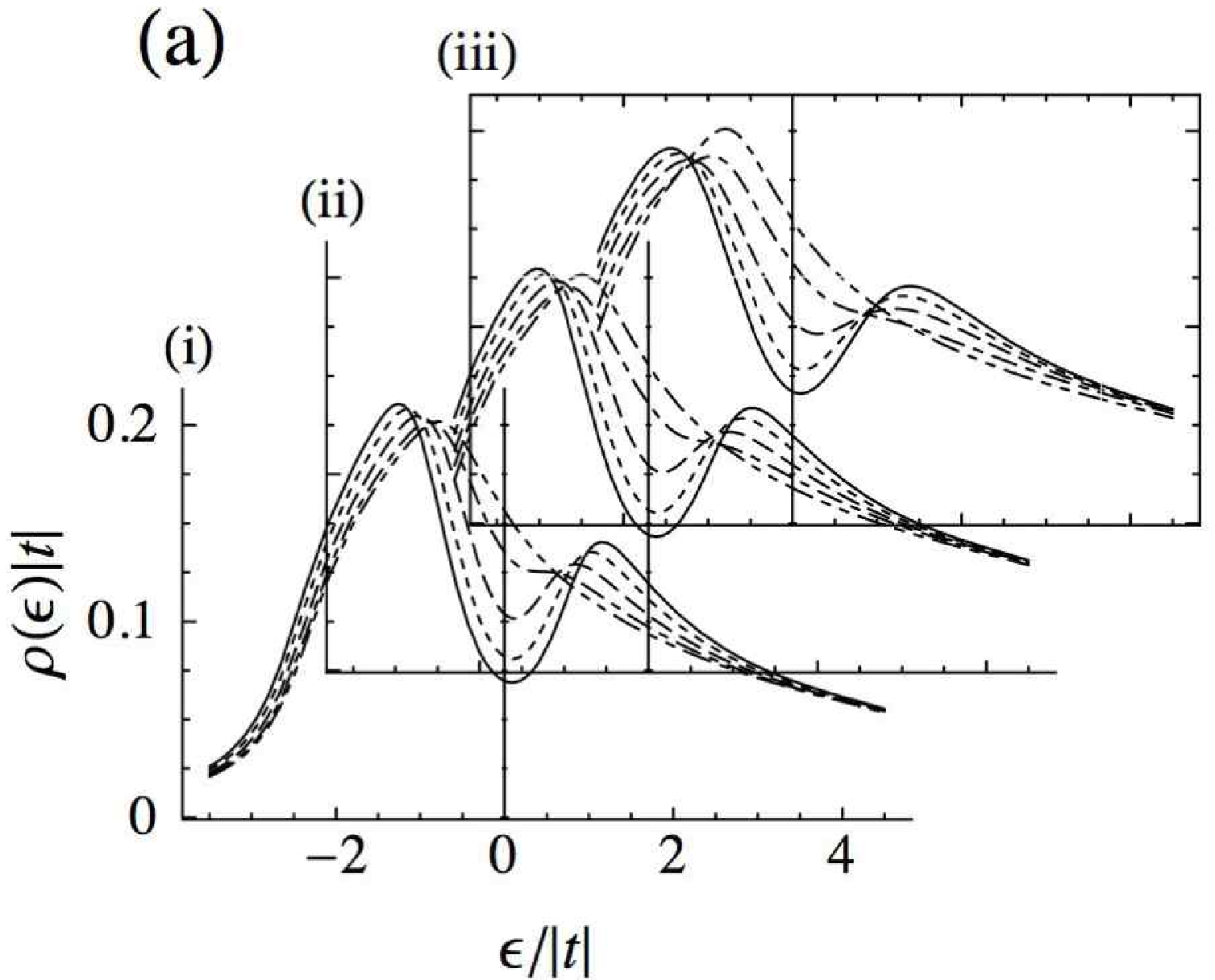}
\includegraphics[width=7.0cm]{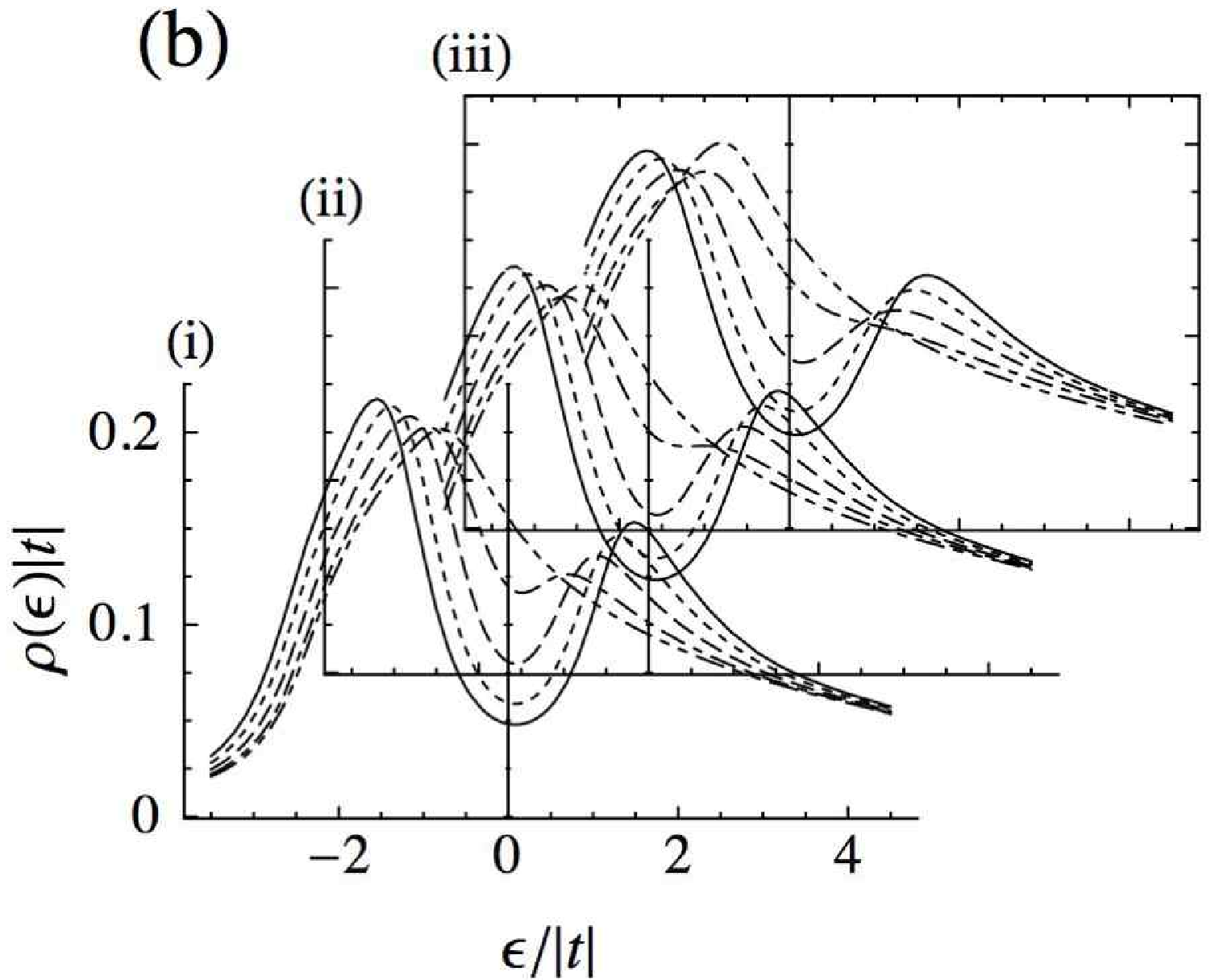}
\includegraphics[width=7.0cm]{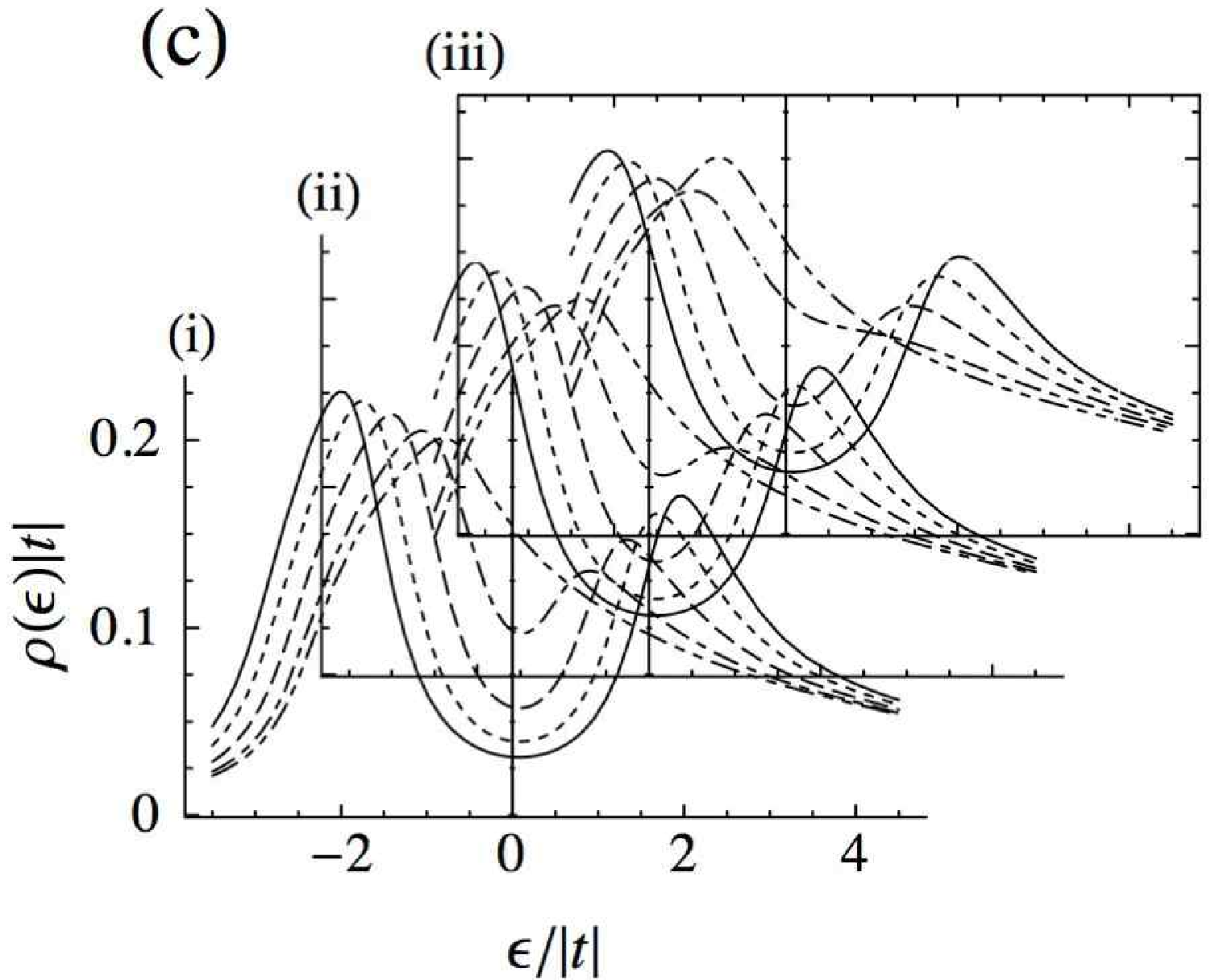}
\caption[3]{
Density of states $\rho(\varepsilon)$
for the $s$ wave,  $\kappa=0$, 
$g_{s}=4$, $\gamma/|t|=0.5$, and $\mu/|t|=-0.5$: 
(a) $k_BT_c/|t|=0.1$, (b) $k_BT_c/|t|=0.2$, 
and (c) $k_BT_c/|t|=0.4$;
 (i) $\Gamma_{SC}=0.1$, (ii) $\Gamma_{SC}=0.3$,
and (iii) $\Gamma_{SC}=1$.
In each figure, solid, dashed, broken, chain, 
and chain double-dashed lines show
$\rho(\varepsilon)$ for $\delta=0.01$, 0.03, 0.1, 0.3, and 1,
respectively. A pseudogap structure is more prominent for higher  $T_c$, smaller $\delta$,
and smaller $\Gamma_{SC}$. No prominent
one is present in any spectrum for the isotropic case $(\delta=1)$.
}
\label{rho-s}
\end{figure*}
\begin{figure}
\hspace*{-0.5cm}
\includegraphics[width=7.0cm]{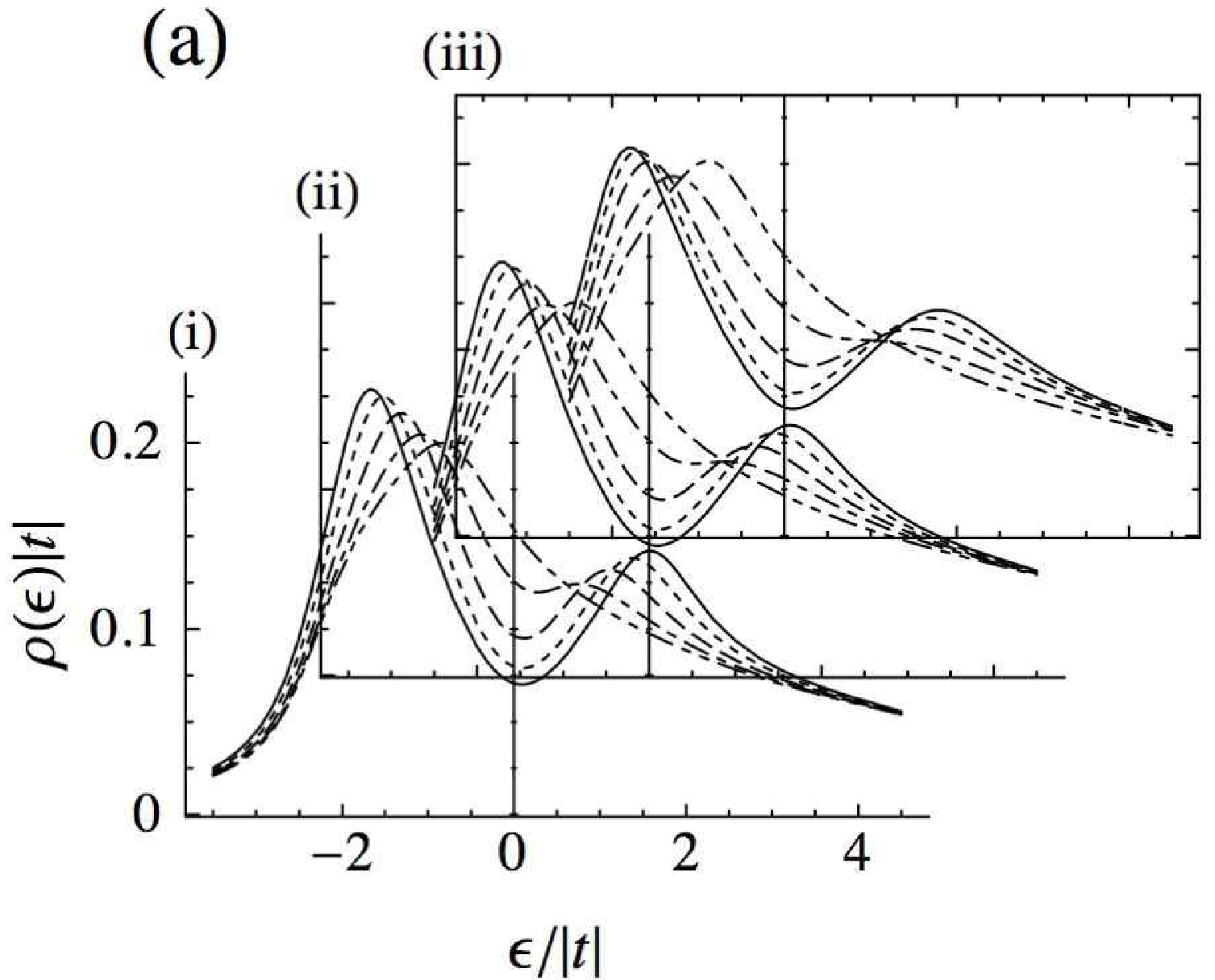}
\includegraphics[width=7.0cm]{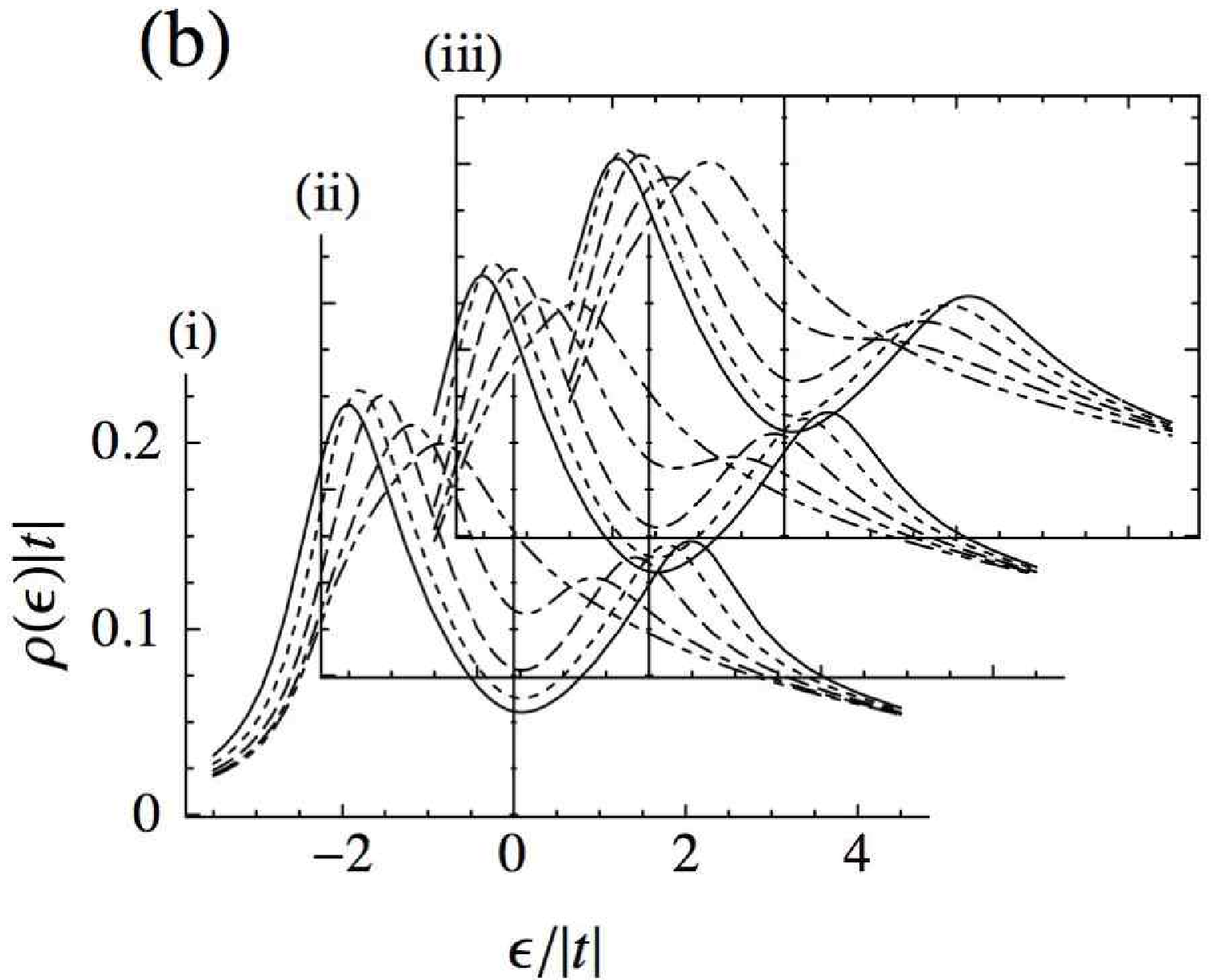}
\includegraphics[width=7.0cm]{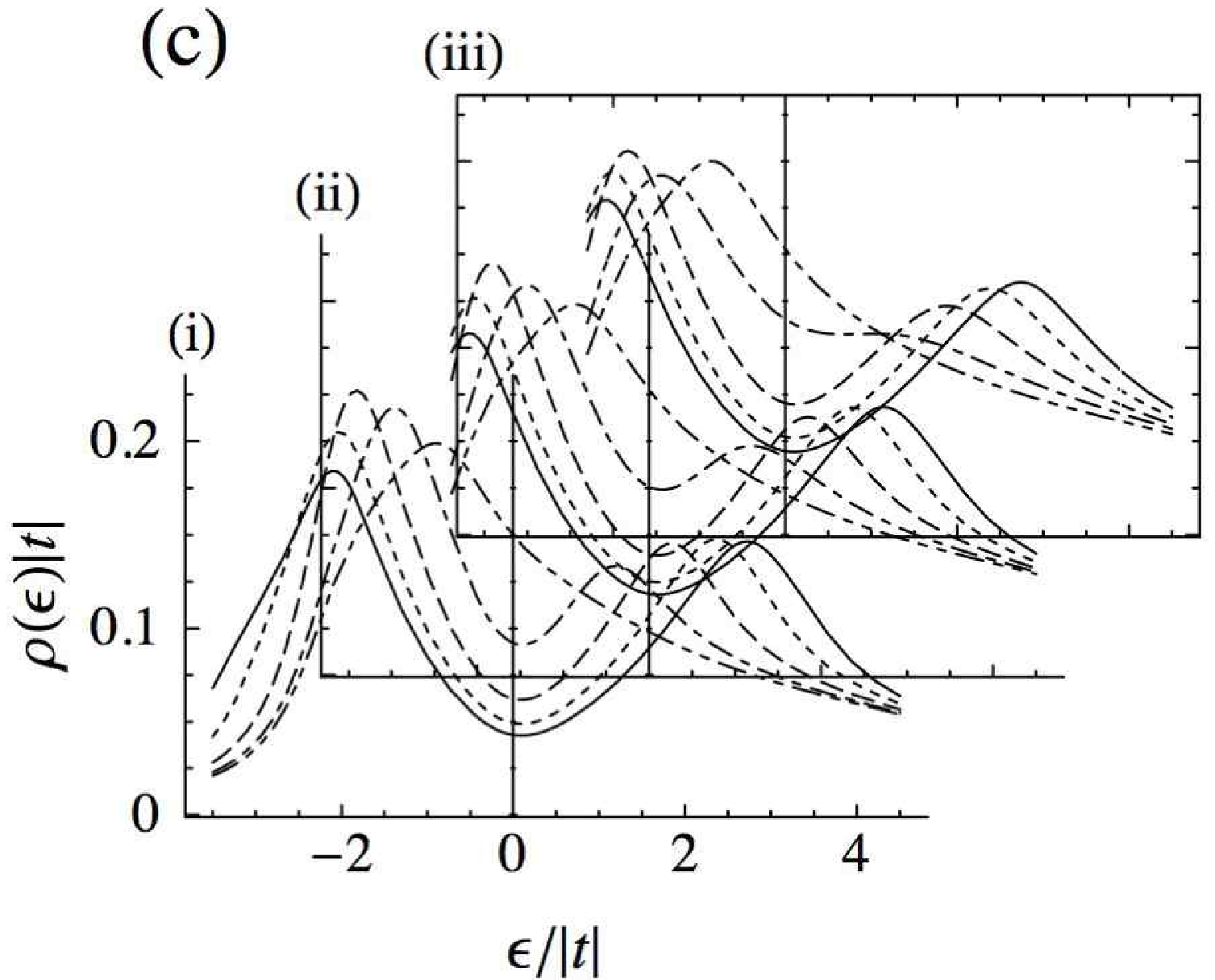}
\caption[4]{
Density of states 
$\rho(\varepsilon)$ for the $d\gamma$ wave,
$\kappa=0$,  $g_{d\gamma}=4$, 
$\gamma/|t|=0.5$, and $\mu/|t|=-0.5$.
See  also the figure caption of Fig.~\ref{rho-s}. 
No essential difference can be seen between Fig.~\ref{rho-s} for the $s$ wave
and this figure for the $d\gamma$ wave.
}
\label{rho-d}
\end{figure}

Figures~\ref{rho-s} and \ref{rho-d} show the density of states
$\rho(\varepsilon)$
for $\kappa=0$ or at the critical point $T=T_c$.
Large anisotropy of
critical fluctuations or a small $\delta$ such as 
$\delta < 0.1$-0.3 is indispensable for the opening of
a prominent pseudogap at $T_c$.
Smaller energy scales $\Gamma_{SC}$ or $\Gamma_{SC}|t|$ 
are favorable for the opening of pseudogaps.
A pseudogap is  more prominent for a higher  $T_c$.
Since higher $T_c$ are mainly caused by larger $g_\Gamma$,
this tendency 
must be  larger when the dependence of $T_c$ on $g_\Gamma$
is considered than it is  in Figs.~\ref{rho-s} and \ref{rho-d}. 
In the isotropic case $(\delta=1)$, on the other hand,
a pseudogap structure is  absent or  subtle at $T_c$.

Since a pseudogap opens  because quasiparticle spectra
around the chemical potential are swept away
due to large lifetime widths,  
 the size of the pseudogap is mainly determined from the  peak width 
of the imaginary part of the selfenergy.  
The peak width is about $2|t|$ for  parameters considered
in this paper so that the size  is also about
$2|t|$; the  peak width and the size are larger for a higher $T_c$ and a smaller $\delta$.

 Although spectra of $\rho(\varepsilon)$ are slightly 
different between the $s$-wave case shown in Fig.~\ref{rho-s} and 
the $d\gamma$-wave case shown in  Fig.~\ref{rho-d}, there is no
essential difference between them as long as $U_{s} \simeq U_{d\gamma}$
or $g_{s} \simeq g_{d\gamma}$.
The anisotropy of critical fluctuations
within planes, the $s$ or $d$ wave, plays a minor role
in the opening of pseudogaps.

\section{Application to cuprate-oxide superconductors }
\label{SecApplication}

When cuprate oxides  are considered,
the formulation presented in Sec.~\ref{SecFormulation}  should be extended to 
treat a  repulsive  strong-coupling regime, 
$U_{0}/|t| \agt 8$  and $|U_1/t|\ll 1$;
we should use
the so called $d$-$p$ or the $t$-$J$ model. \cite{ZhangRice}
The most serious issues are what are single-particle elementary excitations 
or quasiparticles, which are bound into Cooper pairs,
and what is an effective attractive interaction, which 
works between the quasiparticles.
These two issues can be solved by a Kondo-lattice theory, as is 
argued in Appendix.  We present an alternative 
physical argument on these issues  first in this section.

{\it Normal states above $T_c$} are unconventional in the so called
under-doped region, as is discussed in Introduction.  
However, it is certain that the {\it  normal states} 
are Landau's  normal Fermi liquids at least  in the so called over doped region. 
No phase transition is observed within  the {\it  normal states} 
above $T_c$ as a function of doping
concentrations.
According to the 
analytical continuation \cite{AndersonText} as a function of doping
concentrations, therefore,  the
{\it normal states}  above $T_c$ must also  be Landau's  normal Fermi liquids
in the whole metallic region 
even for the so called under-doped region.
The specific heat coefficient  of the so called optimal-doped
cuprate oxides is as large as 
$14$~mJ/K$^2$mol. \cite{gamma1}
Then, we can argue with the use of the Fermi-liquid relation
\cite{Luttinger1,Luttinger2}  that
the  bandwidth of quasiparticles is as small as 0.3~eV or $|t|\simeq 0.04$~eV.
Although the quasiparticle states are often called  mid-gap states,
they are what are predicted by Gutzwiller's theory;
\cite{Gutzwiller1,Gutzwiller2} we call the
quasiparticles Gutzwiller's quasiparticles  in this paper.
An intersite magnetic exchange interaction  can be 
an attractive interaction to form Cooper-pairs. \cite{Hirsch}
The main part of the exchange interaction 
in cuprate oxides is the superexchange interaction.
It  is antiferromagnetic and is as large as $J_s =-0.15$~eV.
It has already been shown in 1987
that high-$T_c$ superconductivity can occur
when  Gutzwiller's quasiparticles
are bound into $d\gamma$-wave Cooper pairs due to the superexchange interaction.
 \cite{highTc}  
According to this scenario, 
high-$T_c$ superconductivity 
occurs in an attractive intermediate-coupling regime
$|J_s/t| \simeq 4$ for superconductivity,
which is realized 
in the repulsive strong-coupling regime for electron correlations.
The Kondo-lattice theory, which is briefly argued in Appendix,
is consistent with this physical argument based on the analytical continuation.

Since $k_BT_c/|t|=0.2$ corresponds to $T_c\simeq 100$~K 
and  $\delta$ must be as small as $\delta \alt 0.1$ in cuprate oxides,
Fig.~\ref{rho-d}(b) implies that the opening of a pseudogap at 
$T_c$ must be mainly due to SC critical thermal fluctuations.
It is plausible that 
even if other mechanisms  work 
the fluctuations play a major role in the opening
of  pseudogaps, at least, at $T_c$ and in SC critical regions
of high-$T_c$ cuprate-oxide superconductors.

\begin{figure*}
\hspace*{-0.5cm}
\includegraphics[width=7.0cm]{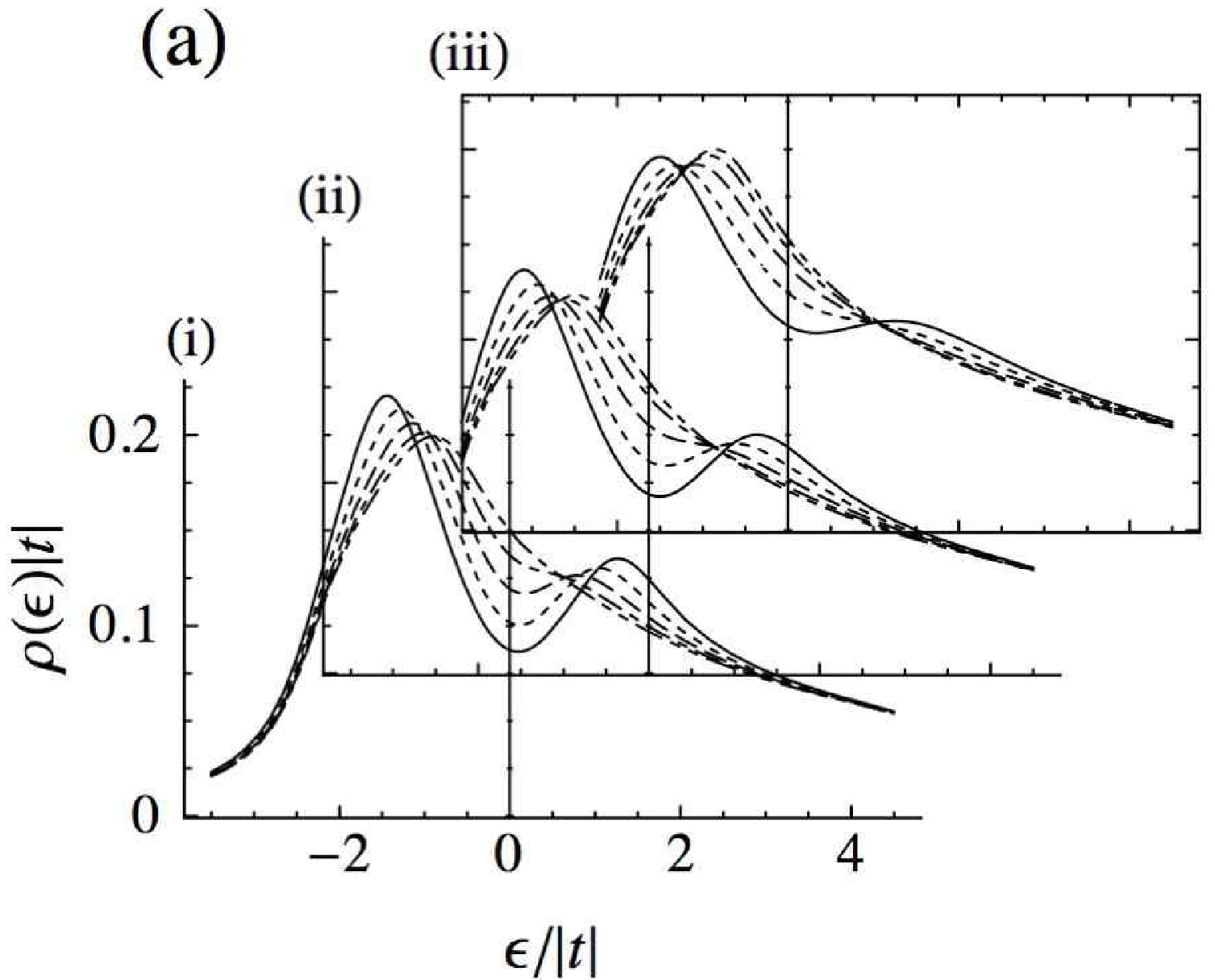}
\includegraphics[width=7.0cm]{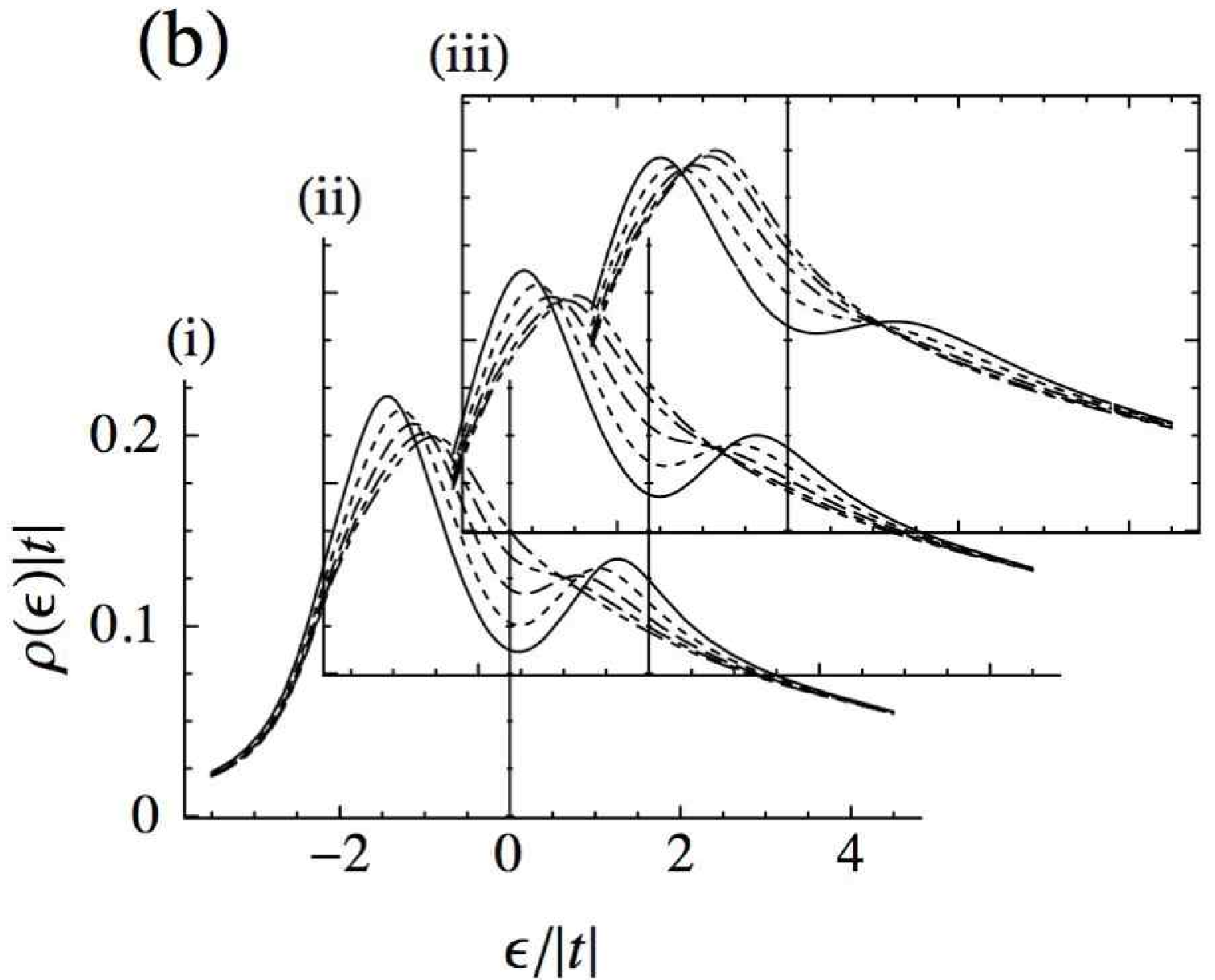}
\includegraphics[width=7.0cm]{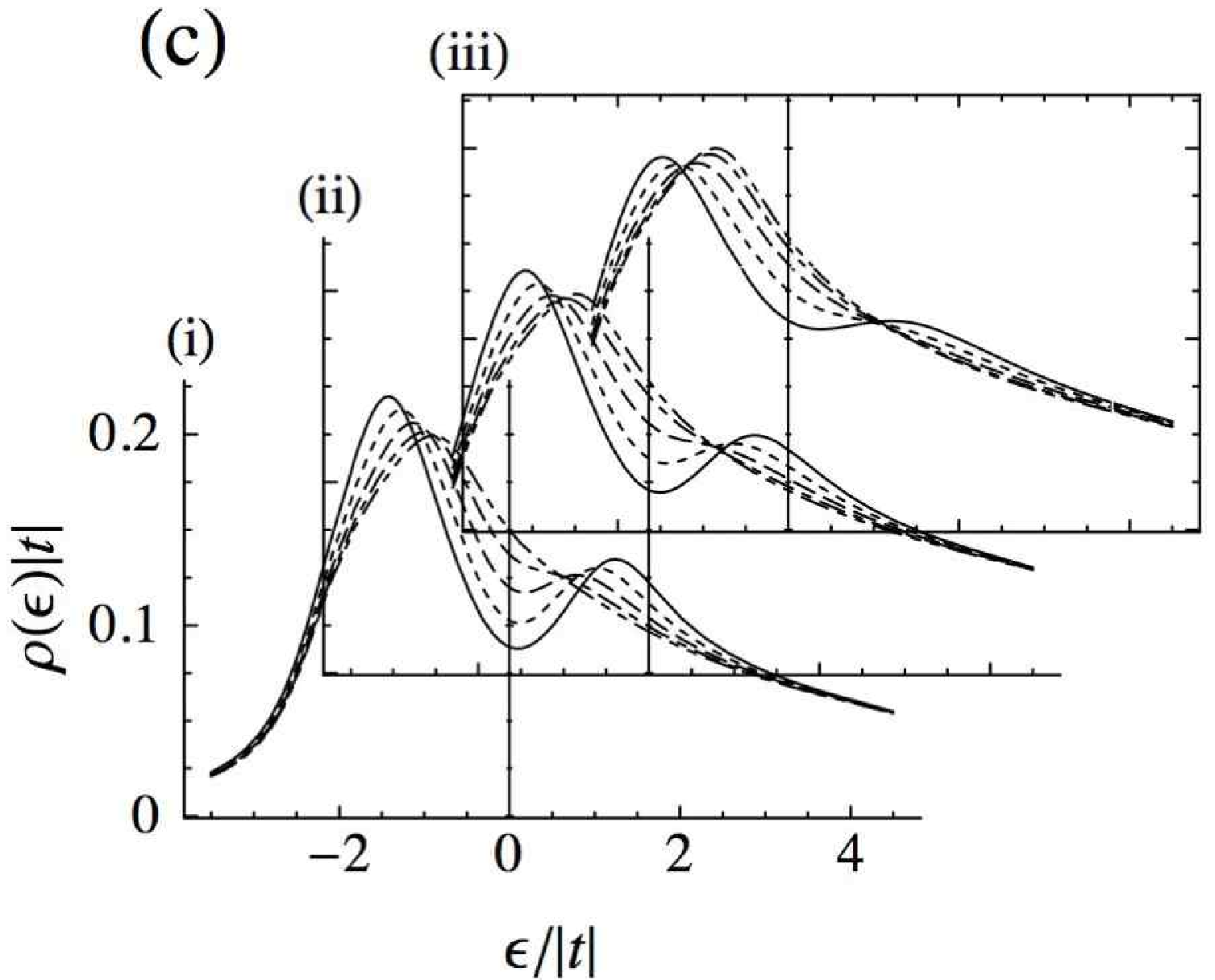}
\includegraphics[width=7.0cm]{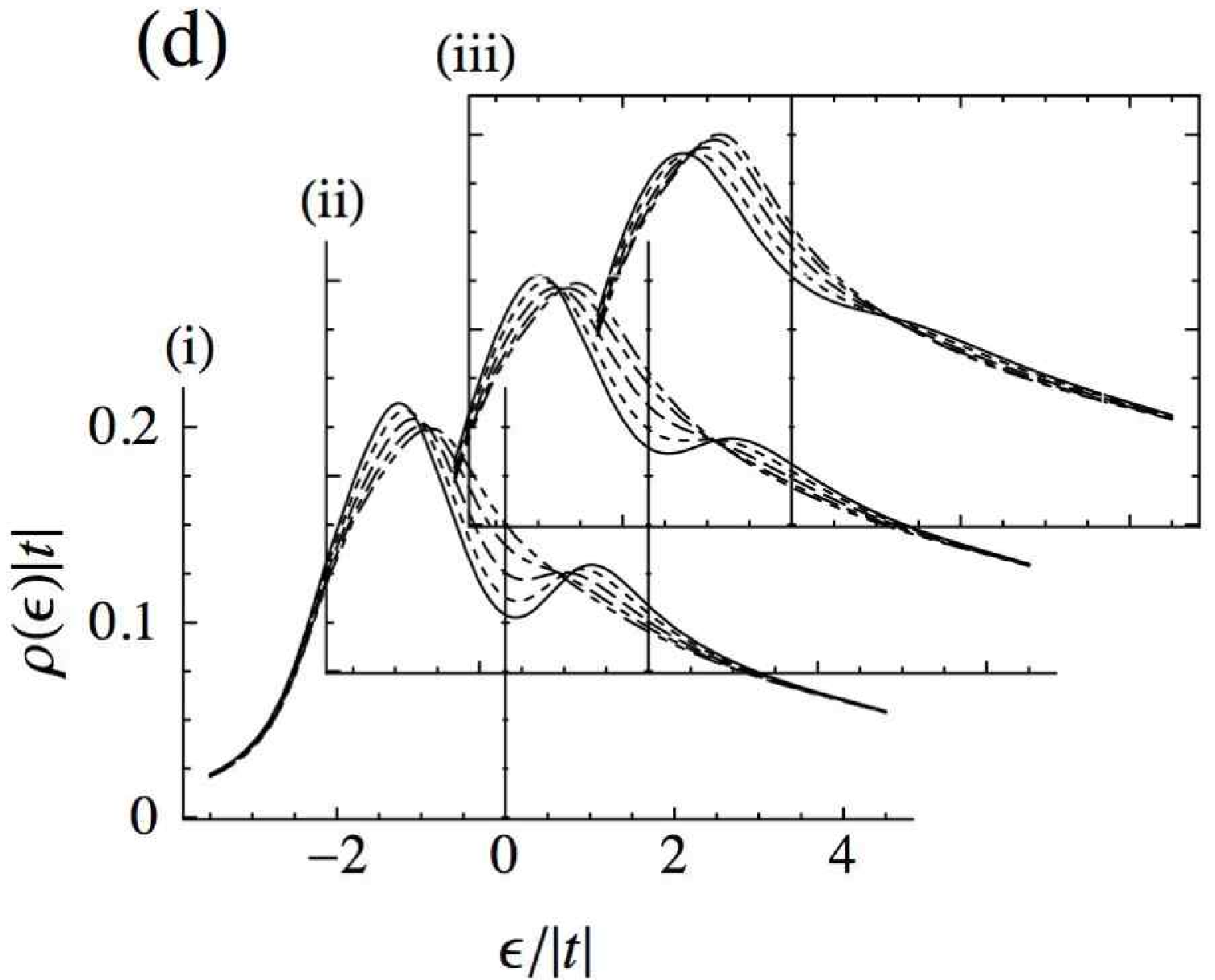}
\caption[5]{
$\rho(\varepsilon)$
at $k_BT/|t|=0.4$ for the $d\gamma$ wave, $g_{d\gamma}=4$, 
$\gamma/|t|=0.5$, and $\mu/|t|=-0.5$:
(a) $\delta=0.01$, (b) $\delta=0.03$, 
(c) $\delta=0.1$, and (d) $\delta=0.3$;
(i) $\Gamma_{SC}=0.1$, (ii) $\Gamma_{SC}=0.3$, and
(iii) $\Gamma_{SC}=1$.
In each figure, 
solid, dashed, broken, chain, and chain double-dashed lines
show $\rho(\varepsilon)$ for $\kappa^2=0.5$,
1, 2, 4, and 8, respectively. 
When either or both of $\kappa^2$ and $\Gamma_{SC}$ are large enough,
a pseudogap structure is absent or subtle. 
}
\label{rho>Tc}
\end{figure*}

If all the parameters such as 
$\kappa$, $\delta$, and $\Gamma_{SC}$,
 were constant as a function of $T$, 
pseudogaps would be developed with increasing $T$, as is shown
 Figs.~\ref{rho-s} and \ref{rho-d}.
Experimentally, however, pseudogaps close at high enough $T$.
It is likely that the temperature dependences of 
$\kappa$, $\delta$, and $\Gamma_{SC}$ are responsible for
the closing of  pseudogaps, for example, 
at $k_BT/|t|\simeq 0.4$ or $T\simeq 200$~K.
Then, we examine what  conditions are needed for the parameters
to exhibit  that a
pseudogap that opens at $k_BT/|t|=0.2$ closes at $k_BT/|t|=0.4$.
It is obvious that $\kappa^2=0$ at $T=T_c$ and
$\kappa^2>0$ at $T>T_c$ or  that $\kappa^2$ increases with
increasing $T$;  $\chi_{\Gamma}(0)\kappa^2$ is almost
constant, as is shown in Eq.~(\ref{EqO1}). It is also obvious that $\Gamma_{SC}$
also increases with increasing $T$.
Figure~\ref{rho>Tc} shows that
when either or both of $\kappa^2$ and $\Gamma_{SC}$ are large enough
 no prominent pseudogap can be seen at $k_BT/|t|=0.4$.
It is interesting to complete the
selfconsistent procedure, \cite{selfconsistency} where
the SC susceptibility is microscopically calculated
and Eq.~(\ref{EqSigma}) is used instead of Eq.~(\ref{EqSigma1}), 
in order to confirm whether or not such temperature dependences
of $\kappa$ and $\Gamma_{SC}$ can be actually reproduced.

\section{Discussion}\label{SecDiscussion}

It is desirable that
the theoretical framework of this paper should be selfconsistently
completed.  However, the selfconsistent procedure
depends on microscopic physical processes or
on what effective Hamiltonian is used,
an intermediate-coupling attractive model 
or a strong-coupling repulsive model. \cite{selfconsistency}
One of the reasons why
we take a phenomenological treatment in this paper
is to demonstrate the essence of the proposed  mechanism 
on pseudogaps due to thermal SC critical fluctuations,
which does not depend on microscopic models.

According to the mean-field theory for $d\gamma$-wave superconductivity, \cite{highTc} 
\begin{equation}
\varepsilon_G(0)/k_BT_c \simeq 4.35,   
\end{equation}
with $\varepsilon_G(0)$ the superconducting gap at $T=0$~K.
Since SC thermal fluctuations  vanish at $T=0$~K,
the reduction of $\varepsilon_G(0)$ must be very small.
As is discussed in Introduction, therefore, 
observed  large ratios \cite{Renner,Ido1,Ido2} of  
 \begin{equation}
 \varepsilon_G(0)/k_BT_c \agt 8 ,
 \end{equation}
are pieces of evidence that $T_c$ are actually reduced by 
the thermal fluctuations, at least, in optimal-doped or
moderately under-doped  cuprate oxides, where $T_c$ are rather high.  
Critical thermal fluctuations must play a major role in the opening of pseudogaps
in such cuprate oxide superconductors with rather high $T_c$.

Since SC thermal fluctuations vanish at $T=0~$K, 
we expect that a pseudogap due to the thermal fluctuations
is closing as $T \rightarrow T_c$ in complete two dimensions,
 where $T_c$ is definitely zero.
A similar argument applies to quasi-two dimensions, where $T_c$ can be nonzero.
A SC gap starts to open at nonzero $T_c$ and
a pseudogap starts to open at $T$ a little higher than $T_c$.
When $T_c$ is low but $\varepsilon_G(0)/k_BT_c$ is large,  it is plausible that 
a pseudogap  opens at rather high temperatures and it is
closing as $T$ approaches the low $T_c$.
The pseudogap never smoothly evolves into the SC gap.
It is interesting to examine whether or not 
pseudogaps are actually closing as $T\rightarrow T_c+0$
in  under-doped cuprate oxides.

Critical thermal fluctuations cannot play any significant role
in the opening of a pseudogap in an
almost isotropic three-dimensional superconductor, 
even if it is of an intermediate coupling
for superconductivity so that its $T_c$ is high.
If a prominent pseudogap opens in an almost isotropic superconductor,
a  mechanism or  mechanisms different from the one proposed in this paper
must be responsible for the opening of the pseudogap.

Mercury-based cuprate oxides show very high $T_c$ under pressures.
\cite{ott,chu}
Pressures must reduce the anisotropy so that
the reduction of $T_c$ becomes smaller with increasing pressures.
It is interesting to search for almost isotropic cuprate oxide 
superconductors  with no prominent pseudogap.  
Since the reduction of $T_c$ by critical fluctuations is small,
their $T_c$ can be higher than $T_c$ of quasi-two dimensional ones.
A simple argument implies that
if $\varepsilon_G(0)/k_BT_c=4-5$ are realized
$T_c$ can exceed 200~K.

Transition-metal dichalcogenide and organic 
superconductors  are  also low dimensional 
superconductors. \cite{layer2,layer3} If $T_c$ are high enough and 
$\varepsilon_G(0)/k_BT_c$ are large enough, pseudogaps
must also open in critical regions. 

The opening of pseudogaps is also expected in quasi-one dimension.
 It is  interesting to examine effects of not only thermal fluctuations 
 but also  quantum fluctuations.
 
It is straightforward to extend
the theory of this paper to pseudogaps due to 
spin and charge fluctuations.
When $T_c$ of a spin density wave (SDW) or a charge density wave (CDW)
is  high enough and the anisotropy of SDW or CDW fluctuations is large enough,
a pseudogap must also open in a critical region of SDW or CDW.
Conventional SC fluctuations are developed around the zone center,
while SDW and CDW ones are developed around wave numbers
corresponding to the  nesting of the Fermi surface.
Scatterings by conventional 
SC ones are  forward scatterings so that their contribution to 
the transport relaxation rate  is small, while
those by SDW and CDW ones must contribute to 
the transport  relaxation rate. It is likely that
resistivity is relatively larger in SDW and CDW cases
than it is in conventional SC cases.

Critical temperatures  $T_c$ of under-doped cuprate oxides, which are
close to  an antiferromagnetic insulating  phase, are very low or vanishing.
The vanishment of $T_c$ can never be explained only
in terms of the reduction of $T_c$ due to 
the thermal fluctuations because they  vanish at $T=0$~K.
Other reduction effects of $T_c$ such as those due to disorder,  
SDW or antiferromagnetism, 
and so on have to be considered to explain the vanishment of $T_c$. 

The so called  zero-temperature pseudogap  (ZTPG) is observed 
at very low temperatures in  under-doped cuprates. \cite{hanaguri,Davis}
Thermal critical fluctuations can never explain ZTPG because
their effects are small at low temperatures.
A mechanism of ZTPG  is proposed in a previous paper.\cite{pairdensity} 
As is discussed in Appendix, magnetic exchange interactions
are responsible for superconductivity as well as magnetism 
 in cuprate oxide superconductors.
Then, the competition or an interplay 
between superconductivity and antiferromagnetism or 
SDW can play a crucial role.
The ZTPG phase must be never a normal Fermi-liquid phase, but 
it must be a non-Fermi liquid phase where 
SC and SDW order parameters coexist.
Experimentally, antiferromagnetic spin fluctuations are well developed 
in under-doped cuprates.
Disorder or large lifetime widths of quasiparticles due to disorder
can play a role in the stabilization of SDW.
\cite{KL-theory3} 
The Brillouin zone is folded by the SDW.
Then,  the condensation of Cooper pairs between two quasiparticles 
around one of  edges of the folded Brillouin zone 
or Cooper pairs  whose total momenta are 
$\pm 2m{\bf Q}_{SDW}$, with 
${\bf Q}_{SDW}$ being a wave number of SDW and 
$m$ being an integer such as 
$m=1$, 2, 3, and so on, can occur in addition to
that of conventional Cooper pairs with zero total momenta.
A $4a$-period stripe structure can arise form an $8a$-period
 single-{\bf Q} SDW;
a $4a \times 4a$ checker-board structure can arise from 
a double-${\bf Q}$ SDW; 
a fine structure similar to that of ZTPG  can arise from 
the coexistence of the single-${\bf Q}$ or double-${\bf Q}$ SDW and 
a multi-${\bf Q}$ pair density wave of $d\gamma$-wave Cooper pairs.
\cite{pairdensity} 
On the other hand,  the normal phase above $T_c$
has no phase boundary between under-doped and over-doped regions.
Then, the examination of this paper implies that  
a pseudogap due to thermal SC and SDW critical fluctuations can open
in the normal phase of under-doped cuprates  where
ZTPG and the checker-board structure are present below $T_c$.

\section{Conclusion}
\label{SecConclusion}

We study the role of  the anisotropy of superconducting critical
thermal  fluctuations in the opening of a pseudogap in a quasi-two
dimensional superconductor.
The thermal fluctuations are developed in a critical region 
provided that  the anisotropy is large enough and the
critical region is extended to high enough temperatures. 
A large ratio of $\varepsilon_G(0)/k_BT_c$, with
$\varepsilon_G(0)$ being the superconducting gap at $T=0~$K, 
is a piece of evidence of well developed thermal fluctuations;
thermal fluctuations can reduce $T_c$ 
while they can never reduce $\varepsilon_G(0)$,
which is for $T=0$~K.
The well developed fluctuations make  lifetime
widths of quasiparticles large.
A pseudogap can open because quasiparticle spectra around the
chemical potential are swept away due to the large lifetime widths.
It can open  in a critical region of not only
anisotropic superconductivity such as $d\gamma$-wave 
one but also  isotropic $s$-wave or BCS one. 
Even if $T_c$ is low in a quasi-two dimensional superconductor, 
a pseudogap can also open at temperatures $T$ 
substantially higher than $T_c$
provided that $\varepsilon_G(0)/k_BT_c$ is large enough.
Since thermal fluctuations are vanishing as $T\rightarrow 0$~K, 
the pseudogap of such a low-$T_c$ superconductor
must be closing as $T\rightarrow T_c+0$. 
Since a pseudogap starts to open at  a temperature
higher than $T_c$ while a superconducting  gap starts to open 
just at $T_c$, it never smoothly evolves into the superconducting  gap.
On the other hand, critical thermal  fluctuations cannot cause
the opening of a prominent pseudogap in an almost isotropic three dimensional  
superconductor,  even if its $T_c$ is high.

Superconducting critical thermal fluctuations must play 
a major role in the
opening of pseudogaps  in critical regions of
cuprate-oxide superconductors with
$\varepsilon_G(0)/k_BT_c \agt 8$,
even if other mechanisms work there.
It is interesting to confirm that  pseudogaps above $T_c$ 
never smoothly evolve into  superconducting gaps 
below $T_c$ in cuprate-oxide superconductors.

\section*{Acknowledgments}
The author is thankful for useful discussion to M. Ido, M. Oda, and N. Momono.

\appendix* 
\section{Quasiparticles and attractive interactions in 
cuprate-oxide superconductors}
\label{SecAppendix}

We consider one of 
effective Hamiltonians where the on-site $U$ repulsion plays a crucial role,
such as  the Hubbard model, the periodic Anderson model, 
the $d$-$p$ model, the $t$-$J$ or $t$-$J$-infinite $U$ model,
 \cite{Comt-J}
and so on; the formulation and
argument of this Appendix can be extended almost in parallel to 
 such various on-site $U$ models.
We assume the repulsive strong-coupling regime for electron correlations
or we assume that the on-site $U$ is as large as the bandwidth 
of  unrenormalized electrons or is larger than it.

According to Hubbard's theory,
\cite{Hubbard1,Hubbard2} a band splits into two subbands
called the lower Hubbard band (LHB) and the upper Hubbard band (UHB).  
According to Gutzwiller's theory, \cite{Gutzwiller1,Gutzwiller2}
 with the use of the Fermi-liquid
theory, \cite{Luttinger1,Luttinger2}
a narrow band  of quasiparticles appears around the
chemical potential;  we call them Gutzwiller's band and  Gutzwiller's
quasiparticles. The combination of the two theories implies that the density of
states must be of a three-peak structure, Gutzwiller's band  between 
LHB and UHB.  Actually the formation of Gutzwiller's band at the top
of LHB in less-than-half filling cases is demonstrated in a previous
paper. \cite{slave} 
The Mott-Hubbard splitting occurs in both metallic and insulating phases, 
and Gutzwiller's band or quasiparticles are  responsible for metallic behaviors.

Any mutual interaction
arises from the virtual exchange of bosons, bosonic excitations, or 
bosonic resonance states.
For example, Yukawa's nuclear force arises from that of pions, 
electromagnetic force from that of photons, 
the attractive interaction in conventional BCS superconductors
from that of phonons, and so on. According to this theoretical framework,
the superexchange interaction arises from that
of pair excitations of electrons across LHB and UHB; \cite{ferro}
it  is phenomenologically given in the $t$-$J$ model.
It works even in  metallic phases 
 as long as the the Mott-Hubbard splitting is significant.
The virtual exchange of pair excitations of Gutzwiller's quasiparticles 
plays no role in the arising of the superexchange interaction, but 
another exchange interaction arises from it. \cite{ferro}
Since  the exchange interaction has a novel property that
its strength is proportional to the width of Gutzwiller's band,
we call it a novel exchange interaction in this paper.
Gutzwiller's quasiparticles are bound into Cooper pairs  mainly 
due to the two exchange interactions.
The superexchange interaction is 
antiferromagnetic  for any filling, \cite{ferroJ} so that it is  attractive 
for singlet Cooper pairs such as $d\gamma$-wave ones.
The novel exchange interaction is antiferromagnetic 
when  the chemical potential lies
around the center of Gutzwiller's quasiparticle band,
while it  is  ferromagnetic when 
the chemical potential lies around the upper or lower edge
of Gutzwiller's quasiparticle band.  \cite{ComNovel} 
For singlet Cooper pairs, in general, the novel exchange interaction
is  attractive   in the case of almost half fillings while it is
repulsive  far away from half fillings.

Strong local correlations, which give rise to the three-peak structure,
 and the exchange interactions
can be treated by a Kondo-lattice theory. 
\cite{pseudogap,ferro,KL-theory2,KL-theory3}
The starting point of the Kondo-lattice theory is
the single-site approximation (SSA)
that includes all the single-site terms.
The SSA  is reduce to solving
the Anderson model selfconsistently; \cite{Mapping-1,Mapping-2,Mapping-3}
any single-site term  of the lattice model,
one of  the effective Hamiltonians with large enough on-site $U$, 
is equal to its corresponding term of the Anderson model.
The local Kondo temperature $T_K$ or $k_BT_K$ is defined as
an energy scale of local quantum spin fluctuations of the lattice model.
The three-peak structure corresponds to the so called Kondo peak
between two subpeaks in the Anderson model;
the width of Gutzwiller's band is about $4k_BT_K$.

We  define an intersite exchange interaction $I_s(i\omega_l,{\bf q})$
by following a physical picture for Kondo lattices that
local spin fluctuations at different sites interact with each other by
an intersite exchange interaction:
\begin{equation}\label{EqDefI}
\chi_s(\omega+i0,{\bf q}) = 
\frac{\tilde{\chi}_s(i\omega_l) }
{1 - \frac1{4} I_s(i\omega_l,{\bf q}) \tilde{\chi}_s(i\omega_l)} ,
\end{equation}
with $\chi_s(i\omega_l,{\bf q}) $ 
being the spin susceptibility of the lattice model
and $ \tilde{\chi}_s(i\omega_l)$ being that of the Anderson model.
Gutzwiller's quasiparticles are well defined at $T\alt T_K$.
The main part of the exchange interaction defined by Eq.~(\ref{EqDefI}) 
is composed of the novel exchange interaction $J_Q(i\omega_l,{\bf q})$
in addition to the superexchange interaction $J_s({\bf q})$:
\begin{equation}
I_s(i\omega_l,{\bf q}) = 
J_s({\bf q}) + J_Q(i\omega_l,{\bf q}) - 4\Lambda(i\omega_l,{\bf q}).
\end{equation}
The third term  $- 4\Lambda(i\omega_l,{\bf q})$ includes 
mode-mode coupling terms among intersite spin fluctuations, 
which correspond to those
 considered in the selfconsistent renormalization
(SCR) theory of spin fluctuations. \cite{moriya}
Magnetism at $T\alt T_K$ is characterized as itinerant-electron one.
The  mode-mode coupling term\cite{moriya} or the novel exchange interaction 
\cite{CW-FJO,miyai}
is responsible for the Curie-Weiss (CW) law of itinerant-electron magnetism;
which is responsible for the CW law 
depends on the dispersion relation of quasiparticles. \cite{comCW}
On the other hand, 
Gutzwiller's quasiparticles are never well defined at $T\gg T_K$;
the novel exchange interaction  vanishes. \cite{rkky}
Magnetism at $T\gg T_K$ is characterized as
local-moment one.
The local term $\tilde{\chi}_s(0)$ in Eq.~(\ref{EqDefI})
is responsible for the CW law of local-moment magnetism. 
The Kondo-lattice theory can treat not only itinerant-electron magnetism
but also  local-moment magnetism; it can treat a crossover between them.
\cite{KL-theory3} 

The  selfenergy in the SSA
is expanded at $T\alt T_K$ in such a way that
\begin{equation}\label{EqExpansion}
\tilde{\Sigma}_{\sigma}(i\varepsilon_n) =
\tilde{\Sigma}_{0}+ \left(1-\tilde{\phi}_{\gamma}\right)i\varepsilon_n
+ \left(1-\tilde{\phi}_{s}\right) \frac1{2}\sigma g\mu_B H+ \cdots,
\end{equation}
with $g$ being the $g$ factor, $\mu_B$ the Bohr magneton,
and $H$ an infinitesimally small magnetic field.
According to the Ward relation, \cite{ward}
the irreducible single-site three-point vertex function in spin channels, 
$\tilde{\lambda}_{s}(i \varepsilon_n,i \varepsilon_n + i \omega_l;i \omega_l)$,
is given by \cite{comSame}
\begin{equation}\label{EqThreeL}
\tilde{\lambda}_{s} (i \varepsilon_n,i \varepsilon_n+i \omega_l;i \omega_l)
= \frac{2\tilde{\phi}_{s} }{U \tilde{\chi}_{s}(i \omega_l)}  
\left[1 + O\left( \frac1{U \tilde{\chi}_{s}(i\omega_l)} \right)
\right],
\end{equation}
for $|\varepsilon_n| \rightarrow +0$ and 
$|\omega_l| \rightarrow +0$.
When  we approximately use Eq.~(\ref{EqThreeL}),
with higher-order terms in
$1/U \tilde{\chi}_{s}(i\omega_l)$ being ignored,
for  small enough
$|\varepsilon_n|$ and  $|\omega_l| $,  it follows that
\begin{equation}\label{SpinFluctI}
\frac{1}{4} U^2 
\tilde{\lambda}_{s}^2 (i \varepsilon_n,
i \varepsilon_n \!+\! i \omega_l;i \omega_l)
\left[
\chi_{s}(i \omega_l,{\bf q}) - \tilde{\chi}_{s}(i \omega_l) 
\right] = 
\tilde{\phi}_{s}^2 \frac{1}{4} I_{\rm s}^* (i\omega_l, {\bf q}) 
\end{equation}
with 
\begin{equation}\label{EqEnhanced-I}
I_s^*(i\omega_l,{\bf q}) = 
\frac{I_s(i\omega_l,{\bf q}) }
{1 - \frac1{4} I_s(i\omega_l,{\bf q}) \tilde{\chi}_s(i\omega_l)}  .
\end{equation}
The left-hand side is the mutual interaction due to spin fluctuations;
the single-site term is subtracted  because it
is considered in the SSA. 
The exchange interaction $I_s^*(i\omega_l,{\bf q}) $  is nothing but an 
exchange interaction enhanced  by spin fluctuations.
The right-hand side is that due to the enhanced exchange interaction;
$\tilde{\phi}_{\rm s}$ appear as effective  single-site vertex functions.
The two mechanisms of  attractive interactions to form Cooper pairs,
the  so called spin-fluctuation mechanism and the 
exchange-interaction mechanism, 
are essentially the same as each other. \cite{TwoMechanisms}

A starting or {\it unperturbed} state is
constructed  in the SSA; it is definitely a normal Fermi liquid. Then, 
 intersite effects  can be  perturbatively considered
in terms of $I_s(i\omega_l,{\bf q})$ or $I_s^*(i\omega_l,{\bf q}) $.
Since this formulation, which is a perturbation theory starting 
from the {\it unperturbed} state
constructed in the non-perturbative SSA  theory,
is consistent with the physical picture for Kondo lattices,
we  should call it a Kondo-lattice theory.
Since the SSA  is rigorous for paramagnetic phases with
no order parameter in infinite dimensions, \cite{metzner}
the  perturbative theory can also be formulated as a $1/d$-expansion theory,
with $d$ the spatial dimensionality. 
The SSA is also called the dynamical mean-field theory (DMFT) \cite{georges}
or the dynamical coherent potential approximation (DCPA). \cite{kakehashi}
Leading order effects in $1/d$ are 
not only local correlations considered in the SSA, which correspond to  
dynamical mean fields considered in DMFT,
the dynamical coherent potential considered in DCPA, 
and  single-site terms or  local effects related with them,
but also conventional Weiss's mean fields of certain instabilities.
\cite{leading}    
All the other effects or terms are of higher order in $1/d$.
Not only the Weiss's mean fields but also higher order effects in $1/d$
can be perturbatively considered.

Taking the Kondo-lattice theory, we can develop
a theory of superconductivity occurring
in the vicinity of the Mott-Hubbard transition 
almost in parallel to that of this paper.
\cite{pseudogap,highTc,KL-theory2,KL-theory3}
Effectively or eventually,  $t$ and $t^\prime$ 
of this paper are replaced by those of Gutzwiller's quasiparticles.
The on-site part of the eventual mutual interaction is definitely strongly repulsive.
Therefore, $T_c$ of $s$-wave or BCS superconductivity cannot be high.
On the other hand, it plays no role in the effective coupling constant
of $d\gamma$-wave superconductivity. 
The attractive interaction given by Eq.~(\ref{SpinFluctI}) includes
 not only  the superexchange interaction
and the novel exchange interaction themselves but also their enhanced ones, which
include effects due to the nesting of the Fermi surface,
the so called inter-nodal scatterings in the $d\gamma$-wave case, and so on.
Although it works between not only nearest-neighbor sites but also neighboring sites,
its nearest-neighbor part play a major role, at least, provided that
the system is a little far away from the critical point of antiferromagnetism.
The attractive interaction of this paper
is replaced by the nearest-neighbor component of
$I_s^*(i\omega_l,{\bf q})$; we simply denote an averaged one over
its low-energy  part  by $I_s^*$.
When the {\it unperturbed}  state is
constructed in the SSA,
$U_{d\gamma}$ for the $d\gamma$ wave
is replaced by 
\begin{equation}\label{EqU*-1}
U_{d\gamma} \rightarrow \tilde{U}_{d\gamma}^*=\frac{3}{4}I_s^* \bigl(
\tilde{\phi}_s / \tilde{\phi}_\gamma \bigr)^2,
\end{equation}
where the factor 3 is due to the three spin channels.
However, Gutzwiller's quasiparticles constructed in the SSA are further renormalized by
SC and antiferromagnetic spin fluctuations.
We should use
the mass renormalization factor $\phi_\gamma({\bf k})$,
which includes such an intersite renormalization
in addition to the single-site renormalization, instead of $\tilde{\phi}_\gamma$.
Then, we should take  
\begin{equation}\label{EqU*-2}
 U_{d\gamma}^*=\frac{3}{4}I_s^* \bigl(
\tilde{\phi}_s / \left<\phi_\gamma ({\bf k})\right>\bigr)^2,
\end{equation}
with $\left<\phi_\gamma ({\bf k})\right>$ 
an average over the Fermi surface.

When we consider cuprate-oxide superconductors, we should use
the $d$-$p$ or $t$-$J$ model \cite{ZhangRice} rather than the Hubbard model.
\cite{NotHub}
In the SSA,  it follows that
$\tilde{\phi}_s / \tilde{\phi}_\gamma \simeq 2$
for almost half fillings so that theoretical $T_c$ are too high
to explain observed $T_c$. \cite{highTc}
When  we  use Eq.~(\ref{EqU*-2}) with
$\tilde{\phi}_s /\left<\phi_\gamma ({\bf k})\right> \simeq 0.7\mbox{-}1$ instead of 
Eq.~(\ref{EqU*-1}) with
$\tilde{\phi}_s / \tilde{\phi}_\gamma \simeq 2$,
we can  explain observed $T_c$.

This Appendix can be concluded in the following way.
The {\it unperturbed} state in the Kondo-lattice 
 theory is definitely a normal Fermi liquid.
Since the assumption of the analytical continuation
is nothing but assuming that the {\it normal} state above $T_c$
is a normal Fermi liquid, 
the Kondo-lattice  theory justifies the assumption of 
the analytical continuation  in Sec.~\ref{SecApplication}.
Then, we can examine an instability of the normal Fermi liquid 
or a symmetry breaking such as a superconducting one
caused by conventional Weiss's
mean fields due to the intersite magnetic exchange
interaction $I_s(i\omega_l,{\bf q})$ or $I_s^*(i\omega_l,{\bf q}) $,
which is essentially the same one as the spin-fluctuation mediated
interaction. Experimentally, the exchange interaction is as large as or
 a little smaller than the bandwidth of quasiparticles. Therefore,
high-$T_c$ superconductivity of cuprate oxides 
must occur in the attractive intermediate-coupling regime
for superconductivity, which is realized 
in the repulsive strong-coupling regime for electron correlations,
that is, 
in the vicinity of the Mott metal-insulator transition or crossover.



\begin{thebibliography}{}
\bibitem{selfconsistency}
If the selfconsistency is completed for  a repulsive model that is 
relevant for cuprate-oxide superconductors,
the renormalization of the effective attractive interaction, which is discussed
in Appendix,  has also to be included in the selfconsistent procedure.
\bibitem{ComLambda}
 In the random-phase approximation,
 where the reduction of $T_c$ due to SC critical fluctuations
 is not considered,
 it follows that $\lambda =1$ for the critical exponent.
\bibitem{ComGamma}
 In the limit of $\gamma/|t|\rightarrow +0$,
the formulation of this paper
is reduced to the conventional non-selfconsistent scheme, 
where  bare or unrenormalized Green functions are put into internal electron lines
in Feynman diagrams.
A pseudogap structure is  the sharpest in this limit.
\bibitem{Comt-J}
Any doubly occupied configuration  at a site is excluded in the $t$-$J$ model.
An infinitely large on-site repulsion $U$ enforces this exclusion 
in the $t$-$J$-infinite $U$ model.
\bibitem{ferroJ}
The superexchange interaction can be ferromagnetic in a multi-band model
with a strong enough Hund coupling.
See for example the reference. \cite{ferro}
\bibitem{comCW}
The SCR theory \cite{moriya} considers itinerant-electron magnets
where the density of state is almost constant 
around the chemical potential and the Fermi surface shows no  nesting.
 The mode-mode coupling term is responsible for the CW law
of  such itinerant-electron magnets. \cite{moriya}
On the other hand,
the novel exchange interaction is responsible for the CW  law
of  ferromagnets \cite{miyai} where the density of state has a sharp peak
around the chemical potential
and antiferromagnets \cite{CW-FJO} where the Fermi surface
shows a sharp nesting.
\bibitem{rkky}
 In the periodic Anderson model,
 strongly  hybridized admixtures of almost localized electrons
and conduction electrons are formed into Gutzwiller's quasiparticles 
or the so called heavy electrons
at $T\alt T_K$. On the other hand, they are independent 
compositions at  $T\gg T_K$.  A relevant physical picture for
$T\gg T_K$ is the coexistence of localized spins and conduction electrons;
the Ruderman-Kittel-Kasuya-Yosida (RKKY)
exchange interaction  arises from the virtual exchange of pair excitations 
of conduction electrons, and it  works between localized spins.
Then, we can argue that
a crossover occurs as a function of  $T$  in such a way that
the novel exchange interaction at $T\alt T_K$
turns into  the RKKY exchange interaction at $T\gg T_K$.
The RKKY exchange interaction is never relevant at low temperatures such as
$T\alt T_K$
\bibitem{leading}
Correctly speaking, leading-order intersite effects in $1/d$ are 
Weiss's means fields of CDW and isotopic $s$-wave or BCS
superconductivity in attractive on-site $U$ models and those of
of SDW, ferromagnetism and antiferromagnetism,
in repulsive on-site $U$ models.
Weiss's means fields of other orders 
such as $d\gamma$-wave superconductivity
are of  higher order in $1/d$. 
%
It can be argue that
Weiss's means fields in the presence of  infinitesimally small external fields
are  responsible for the CW law.
The CW law for local-moment magnetism, which is due to 
the temperature dependence of $1/\tilde{\chi}_s(0)$,
is of leading order in $1/d$. 
The CW law for itinerant-electron magnetism due to that  of
the novel exchange interaction is of leading order in $1/d$, but
the CW law due to that of the mode-mode coupling term is of higher order in $1/d$.
\bibitem{NotHub}
According to the conventional derivation,
the superexchange interaction constant between nearest-neighbor sites
is as large as $J_s=-4|t|^2/U$ for the Hubbard model.
This value is significantly reduced when non-zero bandwidths of
LHB and UHB are considered in the virtual exchange process of pair excitations
of electrons across LHB and UHB.
Therefore, the observed $J_s=-0.15$~eV cannot be consistently explained
within the Hubbard model.
\end{thebibliography}
\end{document}